\documentclass[aps,prb,twocolumn,superscriptaddress,floatfix]{revtex4-2}

\usepackage{amsmath}
\usepackage{dsfont} % double slash fonts
\usepackage{siunitx}
\usepackage{lipsum}
\usepackage{float}
\usepackage[caption=false]{subfig}

\usepackage{tikz} 
\usetikzlibrary{shapes,arrows,positioning,backgrounds}
\usetikzlibrary{decorations.pathreplacing,angles,quotes}

\def\iap{Institute of Applied Physics}
\def\acp{Abbe Center of Photonics}
\def\fsu{Friedrich-Schiller-Universit\"at Jena, Germany}
\def\iof{Fraunhofer Institute of Optics and Precision Engineering, Jena, Germany}
\newcommand{\That}[2]{\hat{T}^{#1}_{#2}}
\newcommand{\Rhat}[2]{\hat{R}^{#1}_{#2}}
\newcommand{\id}{\hat{\mathds{I}}}

\begin{document}

% \title{Interaction of reflection paths of light in metasurface stacks}
\title{Equivalence of reflection paths of light and Feynman paths in stacked metasurfaces}

\author{Jan Sperrhake}
\email[]{jan.sperrhake@uni-jena.de}
\author{Matthias Falkner}
\author{Stefan Fasold}
\author{Thomas Kaiser}
\affiliation{\iap, \acp, \fsu}

\author{Thomas Pertsch}
\affiliation{\iap, \acp, \fsu}
\affiliation{\iof}

\date{\today}

\begin{abstract}
We show the existence of virtual polarization states during the interaction of modes in metasurface stacks. In support of our findings we experimentally realize a metasurface stack, consisting of an isotropic layer of nano-patches and an anisotropic layer of nano-wires. Utilizing an analogy to the interaction of electrons at junctions in mesoscopic electron transport via Feynman paths, we present a semi-analytical description of the modal interaction inside this stack. We then derive a series of all possible reflection paths light can take inside the metasurface stack.
% We show the existence of virtual polarization states during the interaction of modes in metasurface stacks. In support of our findings we experimentally realize a metasurface stack, consisting of an isotropic layer of nano-patches and an anisotropic layer of nano-wires. Utilizing an analogy to the interaction of electrons at junctions in mesoscopic electron transport, we present a semi-analytical description of the modal interaction inside this stack. The resulting method transfers the concept of so called Feynman paths from mesoscopic electron transport to the optical physics of metasurface stacks. We then derive a series of all possible reflection paths light can take inside the metasurface stack. This serves as a tool to identify and extract virtual polarization states. Our physical interpretation of these paths in the experimental data of the patch-wire stack leads to an interaction picture of the fundamental modes of the stack. 
\end{abstract}

\pacs{1234}

\maketitle

\section{Introduction} % (fold)
\label{sec:introduction}

The concept of metasurfaces has permeated many aspects of technological advancement in photonics \cite{Rubin2019,Karalis2019,Colburn2018,Zhu2017,Hentschel2017,Aieta2015a,Lin2014}.
Commonly, metasurfaces comprise artificial two-dimensional arrangements of sub-wavelength structures or particles \cite{Genevet2017,Tretyakov}.
They promise arbitrary control of light \cite{Staude2019,Kenanakis2015,Liu2010a} and the creation of precisely engineered photon states \cite{Poddubny2016,Limonov2017,Wang2018,Wang2019}.
Recent examples of metasurface applications, ranging from hyperspectral imaging \cite{Yesilkoy2019,Do2013} to holography \cite{Forouzmand2018,Jin2018,Li2016a}, lensing \cite{Zhou2018a,Arbabi2016a,Kuznetsov2015} and quantum photonics \cite{Wang2018,Wang2019}, substantiated that promise.

Moreover, metasurfaces can explore links between different disciplines of physics, with  recent advances on so called bound states in the continuum as prominent examples \cite{Hsu2013,Monticone2014,Hsu2016,Cerjan2019}.
Similarly, metasurfaces can facilitate the combination of different physical processes in order to gain highly complex optical functionality \cite{Vazquez-guardado2018,Duggan2019}.

% what do I want to emphasize? the difference to 3d structures OR the multifunctionality
% what is multifunctionality??? functionality??? optical response? response function???

Many studies suggest it to be beneficial to combine different metasurfaces in multi-layered stacks \cite{Berkhout2020,Chen2019a,Menzel2015,Zhao2012a}. A recent example enabled
multi-wavelength meta-lensing by combining geometrically independent dielectric metasurfaces \cite{Zhou2018a}. Another work proposed cascading multiple layers of graphene with dielectric spacer layers to create a broadband optical absorber \cite{Lin2019}.

% maybe skip that ------
% Generally, the complex optical functionality of stacks emerges by combining the features of each layer, while having comparatively simple geometries \cite{Sperrhake2019,Yun2018}. In contrast, single-layer approaches usually focus on complex structural designs, being often challenging to model and fabricate \cite{Helgert2011,Hentschel2017}.

% stack citations uising similar structures \cite{Zhao2012a,Menzel2015,Yun2018,Sperrhake2019}

When light propagates through a stack, metasurfaces interact through inter-layer coupling. Adjacent metasurfaces couple either dominantly in the near-field \cite{Zhou2018a,Chen2019a} or in the far-field \cite{Zhao2012a,Sperrhake2019,Berkhout2020}.
The coupling of near-fields depends on the structures of the metasurface and their local wavelength dependent resonances \cite{Limonov2017,Hentschel2017,Vazquez-guardado2018}. Far-field coupling, on the other hand, does not depend on any local resonance of the metasurfaces. Here, the only interaction mechanism between the metasurfaces is of a Fabry-Perot type \cite{Zhao2012a,Yun2018,Sperrhake2019,Berkhout2020}. Due to the resonant characteristic of this mechanism it modifies the far-field coupling of the modes to adjacent metasurfaces \cite{Sperrhake2019,Berkhout2020}. Hence, we call this type of coupling \textit{modal coupling}. Both numerical \cite{Zhao2012a} and semi-analytic \cite{Sperrhake2019} simulations of far-field coupled metasurfaces reveal this phenomenon as part of the overall stack response. However, the actual interaction process during propagation through a stack is irretrievable from the overall response and, thus, remains hidden.

We aim to derive an interaction picture of stacked metasurfaces by expanding the modal interaction into a series of interfering reflected modes. In particular, we explore the interaction of modes inside a stack consisting of both an isotropic layer of gold nano-patches and an anisotropic layer of gold nano-wires. We analyze how isotropic and anisotropic modes contribute to the interaction and how they influence the total response. Finally, we reveal the existence of virtual polarization states during the modal interaction of the stack.

This work was motivated by the concept of electron scattering paths in mesoscopic solid state physics \cite{Datta1995,Robinson1995}. Here, we attempt to compare and partially transfer this concept to the physics of nano-optics in the specific case of metasurface stacks.

The study of conduction in mesoscopic systems uses descriptive concepts equivalent to the aforementioned modal coupling in metasurface stacks \cite{Bttiker1986,Nazarov2015,Shavit2019}. Specifically, the process of electron scattering at junctions in mesoscopic structures can be considered analogously to the scattering of light at nano-structures \cite{Shavit2019,Texier2016,Li1994a}.

% Reference and analogy to solid state physics
Scattering processes can be described by a set of connected ports in or out of which particles or waves can be transmitted or reflected \cite{Bttiker1986,Li1996}. In the case of mesoscopic electron transport this is the interaction of electrons from different leads at a given junction. Whether an electron is transmitted or reflected into a specific port or not is given by a probability \cite{Buttiker1988,Bttiker1986,Bttiker1985}. Thus, for each combination of ports and whether the interaction results in transmission or reflection there exists a certain combined probability. When a scattering process is complete the final path an electron took can be described as a sum of all its possible paths, weighted by their probability for a given initial port \cite{Robinson1995}. 
Therefore, these paths give a picture of the interaction during the scattering process. In electron scattering theory they represent what is sometimes called the 'Feynman paths' of the system \cite{Datta1995,Robinson1995}. 

% Interpreting metasurfaces as junctions of optical modes, we search for an analogy of Feynman paths between stacked metasurfaces.

Similar to the scattering of electrons at junctions the interaction of light with metasurfaces can be formulated as a scattering problem and described by a set of connected ports \cite{Li1996}. Using scattering matrices (S-matrices) these ports describe the transmitted and reflected modes in different diffraction orders and polarization states \cite{Li1996,Menzel2015}. Additionally, the scattering ports encode whether a mode propagates from front to back of a metasurface or vice versa. Here, we focus on the interaction in modally coupled stacks and establish our model based on the fundamental mode approximation (FMA) \cite{Simovski2007,Simovski2007a}. The FMA is valid if the constituent metasurfaces of the stack are homogeneous \cite{Simovski2009} and their separation is large enough such that adjacent metasurfaces only couple via fundamental modes \cite{Menzel2015}.

\section{Theory of stacked metasurfaces in the fundamental mode approximation}
% Theoretical background: FMA and SASA stacking
Homogeneity implies that both the structures of a metasurface and their lateral separation are smaller than the wavelength of incident light. Furthermore, in the case of a metasurface with periodically arranged structures a fundamental mode is given by the zeroth diffraction order for perpendicularly incident light. If coupling is dominated by fundamental modes, higher diffraction orders have decayed evanescently, which is what the FMA implies \cite{Menzel2015}.

In the FMA regime, the metasurfaces of a stack can each be described by four ports, representing transmission and reflection in two directions. An S-matrix $S_i$ representing the $i$th layer then takes the form of a $2\times2$ block matrix of $2\times2$ Jones matrices for reflection and transmission \cite{Li1996,Menzel2015}. The amplitudes of its complex coefficients are the optical equivalents of the scattering probabilities of electrons from different leads at a junction\cite{Bttiker1986,Datta1995}. For light the scattering coefficients additionally distinguish polarization states in a given basis \cite{Menzel2010}.

In order to denote an S-matrix to a complete stack of $N$ layers we can employ Redheffer's star product $.\ast.$ \cite{Redheffer1960} to combine the S-matrices $S_i$ of each layer such that \cite{Li1996,Menzel2015}
\begin{equation}
	S_{\text{stack}}= S_{N} \ast \dots \ast S_i \ast \dots \ast S_1.
\end{equation}
In this notation light propagates along the z-axis from metasurface 1 to $N$. Each occurrence of the star product gives an overlap of the transmission functions of adjacent metasurfaces and includes all contributions of reflections between them. Mathematically, these contributions are represented by a reflection kernel of the form
\begin{equation}
	( \id - \Rhat{\text{b}}{i}\Rhat{\text{f}}{i+1})^{-1},
\end{equation}
marking $2\times2$ matrices with a hat and defining the two-dimensional identity as $\id$. Here, $\Rhat{\text{b}}{i}$ is the Jones-matrix for reflection of layer $i$ when propagating back to front, as referred to by superscript b, and $\Rhat{\text{f}}{i+1}$ of layer $i+1$ when propagating from front to back, as referred to by superscript f.
The reflection kernels contain exactly the Fabry-Perot type interactions of modally coupled metasurfaces.
For a detailed picture of this interaction process it is therefore necessary to decompose it into its individual reflection paths.

\section{Reflection paths in stacked metasurfaces}
\subsection{Geometric expansion of stacked S-matrices}
In the following we will introduce the mathematical approach we employ to find individual reflection paths during the modal interaction between metasurfaces.

We can expand the reflection kernel of the star product of two S-matrices ($N=2$) into a geometric matrix series \footnote{If it is invertible and its block-matrix elements do not take values $>1$. This is usually the case for physical systems including absorption.}, such that
\begin{equation}
\left( \id - \Rhat{\text{b}}{1}\Rhat{\text{f}}{2} \right)^{-1} = \id + \sum_{\alpha=1}^\infty \left( \Rhat{\text{b}}{1}\Rhat{\text{f}}{2} \right)^\alpha.
\end{equation}
Then, each block matrix $\hat{S}^{ij}$ of a stacked S-matrix can be written as a matrix series 
\begin{equation}
	\hat{S}^{ij} = \hat{S}_0^{ij} + \hat{S}_1^{ij} + \hat{S}_2^{ij} + \dots,ª
\end{equation}
where $i,j\in\{1,2\}$ are the S-matrix's block indices.

% Geometric expansion in context
In optics of stratified media such an expansion is generally known as a Bremmer series \cite{Bremmer1951,Li1994a}. It leads to the
optical WKB (Wentzel, Kramers, Brillouin) approximation of the Helmholtz equation for one-dimensionally inhomogeneous media \cite{Bremmer1951}.
For the much more involved case of stacked metasurfaces we separate the response of the stack into a leading order term (i.e. the
WKB term) and a series of consecutive interferometric terms. For two adjacent layers and front to back propagation
this takes the form of
\begin{equation}
\That{\text{f}}{} = \underbrace{\That{\text{f}}{2}\That{\text{f}}{1}}_{\text{leading transmissive term}} + 
\underbrace{\That{\text{f}}{2}\left( \sum^\infty_{\alpha=1}\left( \Rhat{\text{b}}{1}\Rhat{\text{f}}{2} \right)^\alpha 
	\right)\That{\text{f}}{1}}_{\text{interferometric term}}
\label{eq:Tf_geo}
\end{equation}
for transmission and
\begin{equation}
\Rhat{\text{f}}{} = \Rhat{\text{f}}{1} + \That{\text{b}}{1}\Rhat{\text{f}}{2}\That{\text{f}}{1} +  
\That{\text{b}}{1}\Rhat{\text{f}}{2}\left( \sum^\infty_{\alpha=1}\left( \Rhat{\text{b}}{1}\Rhat{\text{f}}{2} \right)^\alpha \right)\That{\text{f}}{1}.
\label{eq:Rf_geo}
\end{equation}
for reflection. The infinite power series of reflection matrices contains all possible paths light can take between layers after consecutive reflections. For coherent excitation these paths will interfere, including the leading transmissive term. However, separating the pure transmission from inter-layer reflections allows us to analyze how these reflection paths influence the final result.

For more than two layers, we need to expand this concept to an arbitrary number of layers.
Using the associativity of the star product \cite{Redheffer1960} eqs. \eqref{eq:Tf_geo} and \eqref{eq:Rf_geo} can be generalized to $N$ layers by applying each new layer to all the previous ones combined. For this, we introduce the multi-index $M_k \stackrel{\text{def}}{=} 1, \dots ,(N-k)$, denoting all modal contributions from the 1st to the $(N-k)$th layer. 
A transmission or reflection matrix equipped with $M_k$ includes all paths and recurring reflections up to the $(N-k)$th layer, \emph{excluding} those from following layers, $N-k+1,N-k+2,\dots$ and so forth. Imagining the propagation through a stack iteratively this is equivalent to successively connecting the scattering ports of each following layer to the input and output ports of all previous layers combined. Obeying the correct propagation directions (forward and backward) we thereby cascade all possible paths through a stack until the final output port to its substrate.

Then, the transmission through an $N$-layer stack can be written as
\begin{align}
\That{\text{f}}{M_k} &= \That{\text{f}}{N-k}
\prod_{p=1}^{N-k-1} 
\left(\id + \sum_{\alpha=1}^{\infty}\left( \Rhat{\text{b}}{M_{\text{p}}}\Rhat{\text{f}}{n_{p-1}} \right)^\alpha \right) \That{\text{f}}{n_{\text{p}}},
\label{eq:Tf_geo_N}
\end{align}
using the compact index notation $n_p \stackrel{\text{def}}{=} N-k-p$.
The occurring reflection matrices can be found recursively. Given the meaning of the multi-index $M_k$ we also have to obey the order of products in eq. \eqref{eq:Tf_geo_N}. In the context of forward propagation, each backward reflection matrix $\Rhat{\text{b}}{M_{\text{p}}}$ has its own frame of reference within the layer system of the stack. This allows us to comprehend where certain bundles of reflection paths originate from, both mathematically and physically.

Generally, a recursive multi-index reflection matrix is determined as follows,
\begin{align}
\Rhat{\text{f}}{M_k} =~& \Rhat{\text{f}}{M_{k+1}}
+ \That{\text{b}}{M_{k+1}}\Rhat{\text{f}}{N-k}\That{\text{f}}{M_{k+1}} \nonumber\\
&+ \That{\text{b}}{M_{k+1}}\Rhat{\text{f}}{N-k} 
\sum_{\alpha=1}^{\infty}\left( \Rhat{\text{b}}{M_{k+1}}\Rhat{\text{f}}{N-k} \right)^\alpha \That{\text{f}}{M_{k+1}}.
\label{eq:Rf_geo_N}
\end{align}
Changing from forward to backward direction simply results in interchanging the superscripts f and b as well as reversing the index order. If $M_k=1$, only the first layer matrices are applied. The case $k=0$ gives the transmission or reflection of the complete system. Note that the order of indices results from applying the matrices right to left. 

\subsection{Interpreting reflection path coefficients}
To gain insight on each single reflection path we can subtract series that are truncated at different orders $\Psi$ \footnote{See supplementary material for information on truncation.}. For brevity, we choose an arbitrary, scalar transmission coefficient $T$ of a stack described by eq. \eqref{eq:Tf_geo_N}. Introducing the subscript notation $\{\Psi\}$ for a series \textit{up to} order $\Psi$, we define
\begin{equation}
T_{\{\Psi\}} \stackrel{\text{def}}{=} \sum_{\alpha=0}^{\Psi} T_\alpha.
\label{eq:uptopsi}
\end{equation}
With this, the $\Psi$th order contribution is given by
\begin{equation}
T_{\Psi} = T_{\{\Psi\}} - T_{\{\Psi-1\}}.
\label{eq:atpsi}
\end{equation}
We call these coefficients \textit{virtual} as they influence the final response of the stack indirectly through interference. Deriving the transmittance of a truncated coefficient $T_{\{\Psi\}}$ yields
\begin{equation}
|T_{\{\Psi\}}|^2 = \sum_{\alpha=0}^{\Psi}|T_\alpha|^2 + 2\sum_{\alpha=1}^{\Psi}\sum_{\beta=0}^{\Psi-\alpha}|T_\beta||T_{\beta+\alpha}|\cos\left(  \delta_{\alpha\beta}\right),
\label{eq:interferometric_intensitiy}
\end{equation}
where the paths of higher order contributions interfere, depending on the phase differences $\delta_{\alpha\beta}=\phi_\beta - \phi_{\beta+\alpha}$ of their respective
phases $\phi_\alpha$.

\begin{figure}
    \centering
    \includegraphics[width=.75\columnwidth]{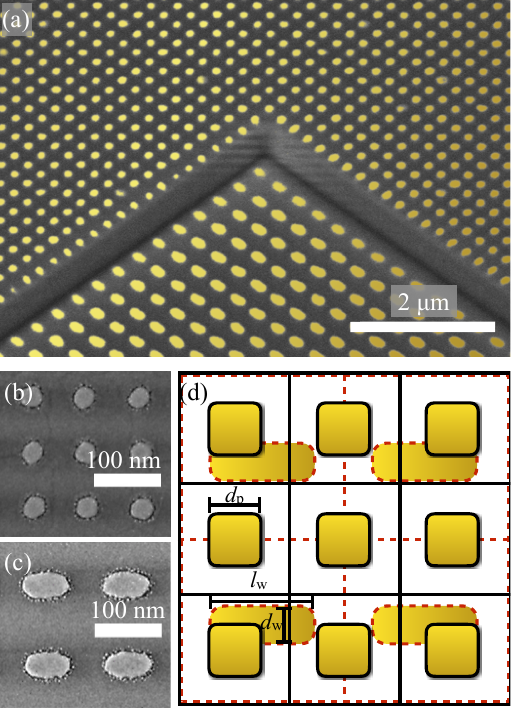}
    \caption{(a) SEM image of the fabricated patch-wire stack revealed by focused ion beam milling. The particles are colored golden for better visibility.
    (b) Single layer field of the upper metasurface with nano-patches.
    (c) Single layer field of the lower metasurface with nano-wires.
    (d) Sketch of superimposed unit cells of the metasurfaces, forming a super-cell of period \SI{600}{nm}. Black lines map unit cells and particles of the upper metasurface and red dashed lines those of the lower metasurface.}
    \label{fig:patch_wire_stack}
\end{figure}

\section{Reflection paths of a patch-wire metasurface stack}
\subsection{Design and fabrication}
% fabrication of a sample
Having established a theoretical framework we now have to ascertain how reflection paths of certain order contribute to an actual physical system.

In order to explore the effect of reflection paths in a real sample we designed and fabricated a metasurface stack consisting of two metasurfaces separated by a glass spacer. The upper or front facing metasurface is comprised of a 2D-array of gold nano-patches with period $\Lambda_{\text{p}}=\SI{200}{nm}$, average diameter $d_{\text{p}}=\SI{70}{nm}$, and height $h_{\text{p}} = \SI{55}{nm}$. The lower metasurface comprises a 2D-array of gold nano-wires with period $\Lambda_{\text{w}} = \SI{300}{nm}$, average lateral dimensions $d_{\text{w}} = \SI{108}{nm}$ and $l_{\text{w}} = \SI{176}{nm}$, and height $h_{\text{w}} = \SI{45}{nm}$. Both metasurfaces were embedded in a glass matrix. Fig. \ref{fig:patch_wire_stack} (a) shows a scanning electron beam (SEM) image of the sample.

Our fabrication technique employed structuring of a two layer resist via electron beam lithography, gold evaporation, and chemical lift-off. 
To obtain reference fields of each metasurface layer in the stack, we fabricated each on two separate fields: the stack itself (fig. \ref{fig:patch_wire_stack} (a)) and a single layer of the respective metasurface (figs. \ref{fig:patch_wire_stack} (b), (c)), resulting in a total of three samples.
After fabricating the first metasurface with this technique, we added a spacer layer using spin-on glass (Futurrex IC1-200) and etched it to the desired thickness of $h_{\text{sp}}=\SI{450}{nm}$. We then fabricated the upper layer metasurface using the same approach as for the lower one. Finally, we added a fused silica cladding layer of thickness $h_\text{c} = \SI{585}{nm}$ by chemical vapor deposition.

\subsection{Semi-analytic modeling }
% modeling
We specifically chose patches and wires for their different symmetry, i.e. $C_4$ and $C_2$, respectively. This gave us the opportunity to analyze the effect of each reflection path on the anisotropic response of the stack, being itself anisotropic with an overall $C_2$ symmetry. Furthermore, the periods of the arrays have a ratio of $\Lambda_{\text{w}}/\Lambda_{\text{p}}=3/2$, creating a super-period of the stacked unit cells, as shown in fig. \ref{fig:patch_wire_stack} (d). Modelling such super-periodic systems usually demands rigorous simulations with high computational effort \cite{Menzel2015}. In our case, however, the spacer thickness of $h_{\text{sp}}=\SI{450}{nm}$ permits applying the FMA, enabling a more efficient semi-analytic approach \cite{Menzel2015,Sperrhake2019}.

We developed a model of the stack utilizing the semi-analytic-stacking algorithm (SASA) presented in \cite{Menzel2015,Sperrhake2019}, which separates the problem into an analytic and a numeric part using S-matrices as described above. Berkhout and Koenderink \cite{Berkhout2020} recently published a comparable approach using transfer matrices. Since we deal with S-matrices and aim to analyze the properties of each scattering channel, SASA is the more suitable choice.

\begin{figure}[ht]
    \centering
    \includegraphics[width=1\columnwidth]{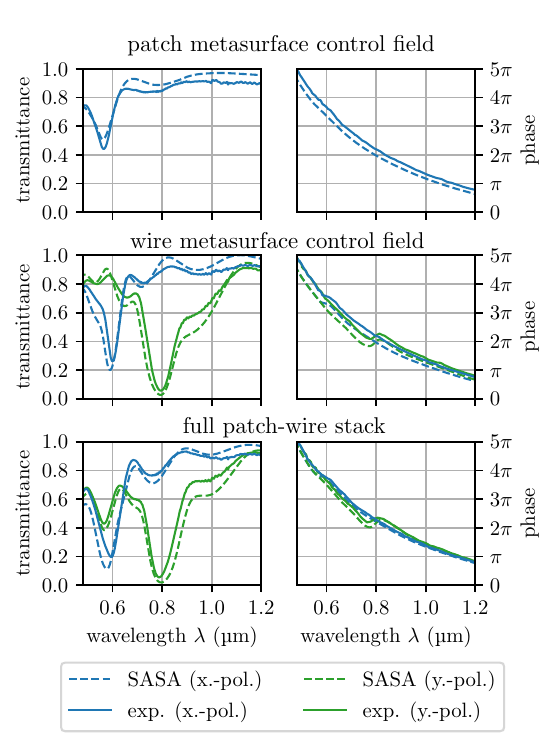}
    \caption{Comparison of measurement and SASA model. The left column of plots shows transmittance and the right column phase. From top to bottom the plots show the results for the single layer control fields of the upper and lower metasurface, and of the full stack at the bottom. Dashed lines refer to SASA results and solid lines to the  measurement. Blue and green differentiate between x- and y-polarization, respectively. Note that only x-polarization is plotted for the patch-metasurface as it is isotropic.}
    \label{fig:patch_wire_subplot}
\end{figure}

First, using the Fourier modal method (FMM) \cite{Li1996,Noponen1994}, we computed the two metasurfaces' S-matrices ($S_{\text{p}}$ for the patches and $S_{\text{w}}$ for the wires) separately for wavelengths ranging from \SI{470}{nm} to \SI{1200}{nm}, while assuming symmetric embedding. Ellipsometric measurements of the materials produced by our fabrication processes supplied refractive index data \cite{Dietrich2012}. Next, all homogeneous dielectric layers, i.e. the spacer, $S_{\text{sp}}$, and the cladding covering the stack, $S_{\text{c}}$, were calculated analytically as propagators of phases \cite{Sperrhake2019}.
Furthermore, we applied Fresnel equations for the interface S-matrix $S_{\text{t}}$ at the top of the stack, representing the glass-air interface of the cladding  \cite{Menzel2015}. In terms of S-matrices the stack is then given by the cascaded star product
\begin{equation}
    S_{\text{stack}} = S_{\text{w}} \ast S_{\text{sp}} \ast S_{\text{p}} \ast S_{\text{c}} \ast S_{\text{t}}.
\end{equation}
The glass wafer at the base of the sample can be considered as a glass half-space with respect to the stack and needs no representation by an S-matrix.

\subsection{Experimental validation}
To ensure the validity of our SASA model we compared it against experimental results.
Using a custom-built in-house characterization setup \cite{Helgert2011,Pshenay-Severin2014a}, we performed interferometric measurements of both the single layer fields and the full stack, simultaneously measuring transmittance and phase in x- and y-polarization.
Fig. \ref{fig:patch_wire_subplot} shows very good agreement between the SASA model and the measurement, both for transmittance and phase.

% define coordinates for tikz objects
\def\yInt{2cm}
\def\yPatch{1cm}
\def\xstep{1cm}
\def\xdist{0.25cm}
\def\xstart{0.5cm}
\begin{figure}[tb]
	\centering
	\begin{tikzpicture}[>=latex]
	
		% draw interface lines
		\draw (0, \yInt)	  -- (2*\xstart + 3*\xstep + 6*\xdist, \yInt) node[right] {$S_{\text{t}}$};
		\draw (0, \yPatch)	-- (2*\xstart + 3*\xstep + 6*\xdist,\yPatch) node[right] {$S_\text{p}$};
		\draw (0,0) 			-- (2*\xstart + 3*\xstep + 6*\xdist,0) node[right] {$S_\text{w}$};
		
		\begin{scope}[thick]
			% draw transmission arrows
			% 0th order
			\draw[->] (\xstart, \yInt)     -- (\xstart, \yPatch);
			\draw[->] (\xstart, \yPatch) -- (\xstart, 0);
			\draw[->] (\xstart, 0) 			 -- (\xstart, -\yPatch) node [below] {$T_0$};
			
			% 1st order
			% set 1
			\draw[->] (\xstart + \xstep, \yInt) -- (\xstart + \xstep, \yPatch);
			\draw[<-, >=right hook] (\xstart + \xstep + \xdist, \yPatch) -- (\xstart + \xstep + \xdist, 1.5*\yPatch);
			\draw[->] (\xstart + \xstep + \xdist, 1.5*\yPatch) -- (\xstart + \xstep + \xdist, \yInt);
			\draw[<-] (\xstart + \xstep + 2*\xdist, \yPatch) -- (\xstart + \xstep + 2*\xdist, 1.5*\yPatch);
			\draw[->,>=left hook] (\xstart + \xstep + 2*\xdist, 1.5*\yPatch) -- (\xstart + \xstep + 2*\xdist, \yInt);
			\draw[->] (\xstart + \xstep + 2*\xdist, \yPatch) -- (\xstart + \xstep + 2*\xdist, 0);
			\draw[->] (\xstart + \xstep + 2*\xdist, 0) -- (\xstart + \xstep + 2*\xdist, -\yPatch) node [below] {$T_1^{(1)}$};
			
			% set 2
			\draw[->] (\xstart + 2*\xstep + 2*\xdist,\yInt) -- (\xstart + 2*\xstep + 2*\xdist, \yPatch);
			\draw[->] (\xstart + 2*\xstep + 2*\xdist,\yPatch) -- (\xstart + 2*\xstep + 2*\xdist, 0);
			\draw[<-, >=right hook] (\xstart + 2*\xstep + 3*\xdist, 0) -- (\xstart + 2*\xstep + 3*\xdist, 0.5*\yPatch);
			\draw[->] (\xstart + 2*\xstep + 3*\xdist, 0.5*\yPatch) -- (\xstart + 2*\xstep + 3*\xdist, \yPatch);
			\draw[<-,>=left hook] (\xstart + 2*\xstep + 4*\xdist, \yPatch) -- (\xstart + 2*\xstep + 4*\xdist, 0.5*\yPatch);
			\draw[->] (\xstart + 2*\xstep + 4*\xdist, 0.5*\yPatch) -- (\xstart + 2*\xstep + 4*\xdist, 0);
			\draw[->] (\xstart + 2*\xstep + 4*\xdist, 0) -- (\xstart + 2*\xstep + 4*\xdist, -\yPatch) node [below] {$T_1^{(2)}$};
			
			% set 3
			\draw[->] (\xstart + 3*\xstep + 4*\xdist,\yInt) -- (\xstart + 3*\xstep + 4*\xdist, \yPatch);
			\draw[->] (\xstart + 3*\xstep + 4*\xdist,\yPatch) -- (\xstart + 3*\xstep + 4*\xdist, 0);
			\draw[<-, >=right hook] (\xstart + 3*\xstep + 5*\xdist, 0) -- (\xstart + 3*\xstep + 5*\xdist, 0.5*\yPatch);
			\draw[->] (\xstart + 3*\xstep + 5*\xdist, 0.5*\yPatch) -- (\xstart + 3*\xstep + 5*\xdist, \yPatch);
			\draw[->] (\xstart + 3*\xstep + 5*\xdist, \yPatch) -- (\xstart + 3*\xstep + 5*\xdist, \yInt);
			\draw[<-] (\xstart + 3*\xstep + 6*\xdist, \yPatch) -- (\xstart + 3*\xstep + 6*\xdist, 1.5*\yPatch);
			\draw[->,>=left hook] (\xstart + 3*\xstep + 6*\xdist, 1.5*\yPatch) -- (\xstart + 3*\xstep + 6*\xdist, \yInt);
			\draw[->] (\xstart + 3*\xstep + 6*\xdist, \yPatch) -- (\xstart + 3*\xstep + 6*\xdist, 0);
			\draw[->] (\xstart + 3*\xstep + 6*\xdist, 0) -- (\xstart + 3*\xstep + 6*\xdist, -\yPatch) node [below] {$T_1^{(3)}$};
			
			% mark up
			\node (air) at (0, 2.5*\yPatch) [left] {air};
			
			\node (top) at (0, 2*\yPatch) [left] {top interface};
			
			\node (clad) at (0, 1.5*\yPatch) [left] {cladding};
			
			\node (patches) at (0, 1*\yPatch) [left] {patch layer};
			
			\node (spac) at (0, 0.5*\yPatch) [left] {spacer};
			
			\node (wires) at (0, 0) [left] {wire layer};
			
			\node (subs) at (0, -.5*\yPatch) [left] {substrate};
								
		\end{scope}
	\end{tikzpicture}
	
	\caption
	{%
	Illustration of zeroth and first order reflection paths. At zeroth order light simply propagates through the stack without reflection. Starting at first order, paths include reflection kernels shown as pairs of curved arrows in the sketch. These are representations of reflection matrix pairs from eq. \eqref{eq:T_patch_wire} with $\Psi = 1$.%
	\label{fig:drawn_paths}%
	}%
\end{figure}

The isotropic patch-metasurface of the upper layer exhibits a single resonance at approximately \SI{580}{nm}. On the other hand, the $C_2$ symmetric wire-metasurface of the lower layer shows two distinct resonances for different polarization at approximately \SI{600}{nm} and \SI{800}{nm}. The isotropic resonance overlaps with polarization sensitive resonances in the stacked configuration. For x-polarization this results in a broader and more prominent resonance at \SI{600}{nm}. However, in y-polarization the transmittance now shows two resonances. The phase is mainly determined by the collective heights of spacer and cladding. Phase jumps at the resonance positions of the single layers combine in the stack.

\subsection{Reflection path extraction}
Having a valid model of the patch-wire stack we now move on to the extraction and analysis of its reflection paths. For brevity, we focus on forward transmission, i.e. propagation from top to bottom of the stack. The S-matrices of the homogeneous layers $S_{\text{sp}}$ and $S_{\text{c}}$ are diagonal matrices with exponential propagation phase terms of the form ). Here, $n$ is the refractive index of the homogeneous medium, $h$ its thickness, and $k_0$ the vacuum wavenumber.
With this we can write the geometric expansion of the patch-wire stack up to order $\Psi$ as
\begin{align}
\That{\text{f}}{M_0} =~& \That{\text{f}}{\text{w}}\left(\id + \sum_{\alpha=1}^{\Psi}\left( \Rhat{\text{b}}{M_1}\Rhat{\text{f}}{\text{w}} \right)^\alpha \right) \mathcal{P}_{\text{sp}}\That{\text{f}}{\text{p}} \nonumber \\
&\times\left(\id + \sum_{\beta=1}^{\Psi}\left( \mathcal{P}_{\text{c}}\Rhat{\text{b}}{\text{t}}\mathcal{P}_{\text{c}}\Rhat{\text{f}}{\text{p}} \right)^\beta \right)\mathcal{P}_{\text{c}}\That{\text{f}}{\text{t}},
\label{eq:T_patch_wire}
\end{align}
where $\mathcal{P}_{\text{sp}}$ and $\mathcal{P}_{\text{c}}$ denote the propagation coefficients of spacer and cladding, respectively. Reflection and transmission at the glass-air interface of the cladding are given by the Jones matrices $\Rhat{\text{b}}{\text{t}}$ and $\That{\text{f}}{\text{t}}$ of the interface S-matrix $S_\text{t}$ \cite{Sperrhake2019}.

\begin{figure}[tb]
    \centering
    \begin{tikzpicture}
        \node[inner sep = 0pt] at (0,0) {\includegraphics[width=1\columnwidth]{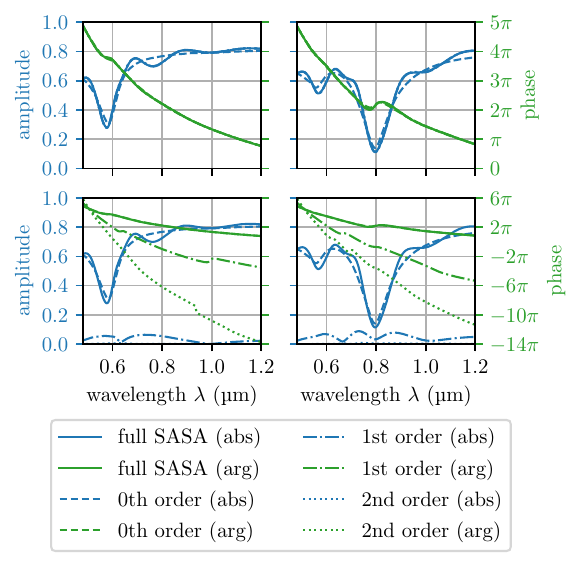}};
        
        % mark up
	    % coordinate grid for support
	    % \draw[step = 0.25cm] (-4,-4) grid (4,4);
	    % \draw[fill=black] (0,0) circle (0.1cm);
        \node at (-4,3.75) {(a)};
        \node at (3.6,3.75) {(b)};
        \node at (-4,1.15) {(c)};
        \node at (3.6,1.15) {(d)};
    \end{tikzpicture}
    \caption{%
	Amplitudes (blue) and phases (green) resulting from a geometric expansion of the SASA model. The top row of plots, (a) and (b), shows transmission $T_{\{\Psi\}}$ when truncating at 0th to 2nd order. The lower row, (c) and (d), shows the separate transmission terms $T_{\Psi}$ contributing at each order $\Psi$. Both cases are compared to the full SASA result without truncation (solid line). The left column, (a) and (c), corresponds to x-polarization. (b) and (d) show y-polarization.%
	}%
    \label{fig:sasa_geo}
\end{figure}
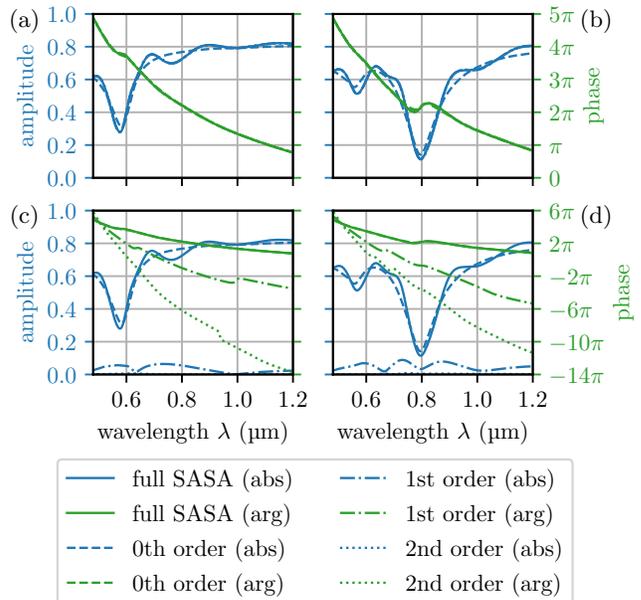

This can be interpreted as follows. The transmission matrices $\That{\text{f}}{\text{t}}$ and $\That{\text{f}}{\text{w}}$ are the input and output ports of the stack. Reading eq. \eqref{eq:T_patch_wire} from right to left, the first parenthesis gives all interactions between the patch-metasurfaces and the top interface of the stack. The second parenthesis includes all interactions between the wire-metasurface and the patch-metasurface. The summations contribute all recurring reflections between those layers. In this context, the multi-index reflection matrix $\Rhat{\text{b}}{M_1}$ bundles all recurring reflections between the top and the spacer of the stack in backward direction.

With the notation from eqs. \eqref{eq:uptopsi} and \eqref{eq:atpsi} as well as using eq. \eqref{eq:Rf_geo_N} to calculate the reflection matrices in \eqref{eq:T_patch_wire}, we can formulate the explicit expressions of the physical reflection paths in eq. \eqref{eq:T_patch_wire}. In $x$-polarization the transmission coefficients of zeroth order and the first three paths contained at first order ($\Psi = 1$) read as
\begin{align}
T_0		  & = T^x_{\text{w}} \mathcal{P}_{\text{sp}} T_{\text{p}} \mathcal{P}_{\text{c}} T_{\text{t}} \label{eq:zerothorder}\\
T_1^{(1)} & = T^x_{\text{w}} \mathcal{P}_{Sp }T_{\text{p}} R_{\text{p}} R_{\text{t}} \mathcal{P}^3_{\text{c}} T_{\text{t}} \label{eq:firstorder1} \\
T_1^{(2)} & = T^x_{\text{w}} R^x_{\text{w}} R_{\text{p}} \mathcal{P}_{\text{sp}}^3 T_{\text{p}} \mathcal{P}_{\text{c}} T_{\text{t}} \label{eq:firstorder2} \\
T_1^{(3)} & = T^x_{\text{w}} R^x_{\text{w}} \mathcal{P}_{\text{sp}}^3 T_{\text{p}}^3 R_{\text{t}} \mathcal{P}^3_{\text{c}} T_{\text{t}}, \label{eq:firstorder3}
\end{align}
where we omitted the superscripts f and b for the sake of readability. Above, eqs. \eqref{eq:firstorder1} through \eqref{eq:firstorder3} show the coefficients of the paths emerging at first order, such that $T_{\{1\}} = T_0 + (T_1^{(1)} + T_1^{(2)} + T_1^{(3)} + \dots)$. The graphical representation of these coefficients in fig. \ref{fig:drawn_paths} shows the paths in the context of the fabricated stack. Whereas the leading transmissive term $T_0$ expresses propagation straight through the stack, the first order paths include different combinations of recurring reflections.

% interpreting the path
The leading transmissive term $T_0$ is composed of the single layer transmission coefficients, with $\mathcal{P}_{\text{c}}$ and $\mathcal{P}_{\text{sp}}$ imposing an additional phase shift. Both the isotropic patch-metasurface and the anisotropic wire-metasurface contribute equally to the combined transmission coefficient. At higher orders each reflection path shows different compositions of the isotropic and anisotropic contributions. Therefore, they add different degrees of anisotropy to the interferometric part of the stack's transmission.

Inputting the SASA results into eqs. \eqref{eq:atpsi} and \eqref{eq:T_patch_wire} we computed both the truncated series $T_{\{\Psi\}}$ and the coefficients $T_\Psi$ of the patch-wire stack numerically. To see how the series converges we truncated this time at second order ($\Psi = 2$). Fig. \ref{fig:sasa_geo} shows amplitude and phase of both sets of coefficients, $\{T_{\{0\}}, T_{\{1\}}, T_{\{2\}}\}$ and $\{T_0, T_1, T_2\}$, both for x- and y-polarized light. Looking at the set of truncated coefficients $T_{\{\Psi\}}$ (figs. \ref{fig:sasa_geo} (a), (b)) we see that the series already approximates the amplitude of the full result well at 1st order. The phase, however, seems to be insensitive to the expansion. But this is no surprise since the phase is mainly determined by the propagation lengths in the stack and its resonances. Any extra phase vanishes due to interference. In contrast, the set of coefficients contributing to each order $T_\Psi$ (figs. \ref{fig:sasa_geo} (c), (d)) shows the accumulated phase of the taken paths, albeit without interference.
Here, we see that the metasurfaces' resonances manifest themselves in the amplitude of the 1st order contribution.

\begin{figure}[tb]
    \centering
    \begin{tikzpicture}
    \node[inner sep = 0] at (0,0) {\includegraphics[width=\columnwidth]{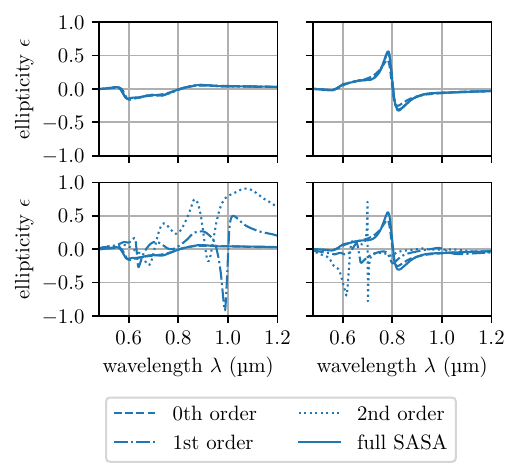}};
    
    % markup
    % \draw[step=0.5] (-4,-4) grid (4,4);
    % \draw[fill=black] (0,0) circle (0.1cm);
    \node at (-2.425,3.275) {(a)};
    \node at (1.15,3.275) {(b)};
    \node at (-2.425,0.6) {(c)};
    \node at (1.15,0.6) {(d)};
    \end{tikzpicture}
    \caption{Ellipticity of transmission coefficients resulting from the geometric expansion of the SASA model. The upper row of plots, (a) and (b), shows the ellipticity $\epsilon$ of transmission coefficients $T_{\{\Psi\}}$ when truncating at 0th to 2nd order. The lower row, (c) and (d), shows the ellipticity of the separate transmission terms $T_\Psi$ contributing at each order $\Psi$. Both cases are compared to the full SASA result without truncation (solid line). The left column, (a) and (c), corresponds to x-polarization. (b) and (d) show y-polarization.}
    \label{fig:elli_geo}
\end{figure}
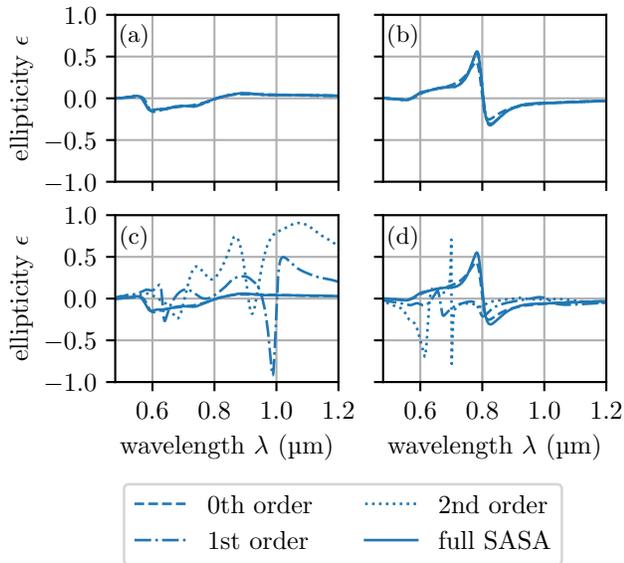

\subsection{Virtual polarization states}
The discussion above showed how and to what degree different orders of reflection paths contribute to the overall stack response. Now, we can pose the question: what other physical insights can we deduce from the properties of reflection paths? Indeed, with an anisotropic stack at hand we can gauge the degree of anisotropy at different expansion orders.

By calculating the ellipticity of the sets of coefficients from fig. \ref{fig:sasa_geo} we can compare the stack's overall anisotropic response to that of each expansion order. From the results shown in fig. \ref{fig:elli_geo} (a) we can conclude that for x-polarization the stack's response is mostly linearly polarized with a slight deviation around the resonance wavelength at \SI{600}{nm}. In y-polarization (fig. \ref{fig:elli_geo} (b)) elliptical polarization is produced around the stronger resonance at \SI{800}{nm}. As before, the geometric series converges already at 1st order.

The ellipticity of the individual reflection paths shows more complex behavior (figs. \ref{fig:elli_geo} (c), (d)). For instance, at first order in x-polarization a circular polarization state emerges at a wavelength of \SI{1000}{nm} (fig. \ref{fig:elli_geo} (c)). We term such states \textit{virtual} polarization states since they interfere with other paths and produce only low amplitudes. This demonstrates that the reflection paths of the patch-wire stack are anisotropic to varying degree. Even though it is small, they have a distinguishable influence on the stacks overall anisotropic response.

\section{Conclusions}
% ... borrowing from electronic transport in mesoscopic systems
In conclusion, we revealed the existence of virtual polarization states of a metasurface stack by analyzing the reflection paths of its internal modal interactions.
Our approach was motivated by the treatment of electron scattering in mesoscopic electron transport using a multiport formalism. This concept is mathematically equivalent to the scattering matrix formalism we employed. Based on this conceptual overlapp we could adopt the analogy of Feynman paths and electron scattering paths to the scattering problem in stacked metasurfaces.

In this work we applied a geometric expansion to the S-matrix of an anisotropic patch-wire metasurface stack under the necessary condition of the FMA. We demonstrated that its transmission could be separated into a leading transmissive term and a series of interferometric terms, representing the reflection paths of the stack. By truncating the series and analyzing its constituent coefficients, we revealed the properties of paths of different order as well as their influence on the overall response.

The knowledge of reflection paths could prove useful in understanding the interaction of more complex stacks with multiple diffraction channels \cite{Chen2019a}. Furthermore, we believe that the concept of Feynman paths could help in developing semi-analytic models of near-field interactions of complex nano-structures which can be challenging to comprehend, even numerically \cite{Helgert2011,Kenanakis2015,Forouzmand2018}.

Finally, we would like to emphasize the benefit of adopting concepts from different fields of physics and identifying their similarities in order to gain more insight on certain physical phenomena.

\begin{acknowledgments}
We would like to thank Prof. Asger Mortensen for discussions leading to the idea of exploiting methods from electronic transport in mesoscopic systems. Furthermore, we would like to thank Prof. Yuri Kivshar for discussions on the scope of the paper. Thanks are also due for or fabrication team, Michael Steinert, Waltraut Gr\"af, Holger Schmidt, and Daniel Voigt, for their support in the experimental realization of the patch-wire stack. We gratefully acknowledge financial support by the German Federal Ministry of Education and Research in the program 'Zwanzig20 - Partnership for Innovation' as part of the research alliance 3Dsensation (grant numbers 03ZZ0466, 03ZZ0471D, and 03ZZ0451 as well as 03Z1H534). Furthermore, we are grateful for funding by the 'Deutsche Forschungsgemeinschaft' (DFG, German Research Foundation) – Project-ID 278747906 and by the European Community within H2020 – Project-ID 899673 (METAFAST).
\end{acknowledgments}

\bibliography{paper_refs.bib}

\begin{thebibliography}{59}%
\makeatletter
\providecommand \@ifxundefined [1]{%
 \@ifx{#1\undefined}
}%
\providecommand \@ifnum [1]{%
 \ifnum #1\expandafter \@firstoftwo
 \else \expandafter \@secondoftwo
 \fi
}%
\providecommand \@ifx [1]{%
 \ifx #1\expandafter \@firstoftwo
 \else \expandafter \@secondoftwo
 \fi
}%
\providecommand \natexlab [1]{#1}%
\providecommand \enquote  [1]{``#1''}%
\providecommand \bibnamefont  [1]{#1}%
\providecommand \bibfnamefont [1]{#1}%
\providecommand \citenamefont [1]{#1}%
\providecommand \href@noop [0]{\@secondoftwo}%
\providecommand \href [0]{\begingroup \@sanitize@url \@href}%
\providecommand \@href[1]{\@@startlink{#1}\@@href}%
\providecommand \@@href[1]{\endgroup#1\@@endlink}%
\providecommand \@sanitize@url [0]{\catcode `\\12\catcode `\$12\catcode
  `\&12\catcode `\#12\catcode `\^12\catcode `\_12\catcode `\%12\relax}%
\providecommand \@@startlink[1]{}%
\providecommand \@@endlink[0]{}%
\providecommand \url  [0]{\begingroup\@sanitize@url \@url }%
\providecommand \@url [1]{\endgroup\@href {#1}{\urlprefix }}%
\providecommand \urlprefix  [0]{URL }%
\providecommand \Eprint [0]{\href }%
\providecommand \doibase [0]{https://doi.org/}%
\providecommand \selectlanguage [0]{\@gobble}%
\providecommand \bibinfo  [0]{\@secondoftwo}%
\providecommand \bibfield  [0]{\@secondoftwo}%
\providecommand \translation [1]{[#1]}%
\providecommand \BibitemOpen [0]{}%
\providecommand \bibitemStop [0]{}%
\providecommand \bibitemNoStop [0]{.\EOS\space}%
\providecommand \EOS [0]{\spacefactor3000\relax}%
\providecommand \BibitemShut  [1]{\csname bibitem#1\endcsname}%
\let\auto@bib@innerbib\@empty
%</preamble>
\bibitem [{\citenamefont {Rubin}\ \emph {et~al.}(2019)\citenamefont {Rubin},
  \citenamefont {D'Aversa}, \citenamefont {Chevalier}, \citenamefont {Shi},
  \citenamefont {Chen},\ and\ \citenamefont {Capasso}}]{Rubin2019}%
  \BibitemOpen
  \bibfield  {author} {\bibinfo {author} {\bibfnamefont {N.~A.}\ \bibnamefont
  {Rubin}}, \bibinfo {author} {\bibfnamefont {G.}~\bibnamefont {D'Aversa}},
  \bibinfo {author} {\bibfnamefont {P.}~\bibnamefont {Chevalier}}, \bibinfo
  {author} {\bibfnamefont {Z.}~\bibnamefont {Shi}}, \bibinfo {author}
  {\bibfnamefont {W.~T.}\ \bibnamefont {Chen}},\ and\ \bibinfo {author}
  {\bibfnamefont {F.}~\bibnamefont {Capasso}},\ }\bibfield  {title} {\bibinfo
  {title} {{Matrix Fourier optics enables a compact full-Stokes polarization
  camera}},\ }\href {https://doi.org/10.1126/science.aax1839} {\bibfield
  {journal} {\bibinfo  {journal} {Science}\ }\textbf {\bibinfo {volume}
  {365}},\ \bibinfo {pages} {eaax1839} (\bibinfo {year} {2019})}\BibitemShut
  {NoStop}%
\bibitem [{\citenamefont {Karalis}\ and\ \citenamefont
  {Joannopoulos}(2019)}]{Karalis2019}%
  \BibitemOpen
  \bibfield  {author} {\bibinfo {author} {\bibfnamefont {A.}~\bibnamefont
  {Karalis}}\ and\ \bibinfo {author} {\bibfnamefont {J.~D.}\ \bibnamefont
  {Joannopoulos}},\ }\bibfield  {title} {\bibinfo {title} {{Plasmonic
  Metasurface “Bullets” and other “Moving Objects”: Spatiotemporal
  Dispersion Cancellation for Linear Passive Subwavelength Slow Light}},\
  }\href {https://doi.org/10.1103/PhysRevLett.123.067403} {\bibfield  {journal}
  {\bibinfo  {journal} {Physical Review Letters}\ }\textbf {\bibinfo {volume}
  {123}},\ \bibinfo {pages} {067403} (\bibinfo {year} {2019})}\BibitemShut
  {NoStop}%
\bibitem [{\citenamefont {Colburn}\ \emph {et~al.}(2018)\citenamefont
  {Colburn}, \citenamefont {Zhan},\ and\ \citenamefont
  {Majumdar}}]{Colburn2018}%
  \BibitemOpen
  \bibfield  {author} {\bibinfo {author} {\bibfnamefont {S.}~\bibnamefont
  {Colburn}}, \bibinfo {author} {\bibfnamefont {A.}~\bibnamefont {Zhan}},\ and\
  \bibinfo {author} {\bibfnamefont {A.}~\bibnamefont {Majumdar}},\ }\bibfield
  {title} {\bibinfo {title} {{Metasurface optics for full-color computational
  imaging}},\ }\href {https://doi.org/10.1126/sciadv.aar2114} {\bibfield
  {journal} {\bibinfo  {journal} {Science Advances}\ }\textbf {\bibinfo
  {volume} {4}},\ \bibinfo {pages} {1} (\bibinfo {year} {2018})}\BibitemShut
  {NoStop}%
\bibitem [{\citenamefont {Zhu}\ \emph {et~al.}(2017)\citenamefont {Zhu},
  \citenamefont {Yan}, \citenamefont {Levy}, \citenamefont {Mortensen},\ and\
  \citenamefont {Kristensen}}]{Zhu2017}%
  \BibitemOpen
  \bibfield  {author} {\bibinfo {author} {\bibfnamefont {X.}~\bibnamefont
  {Zhu}}, \bibinfo {author} {\bibfnamefont {W.}~\bibnamefont {Yan}}, \bibinfo
  {author} {\bibfnamefont {U.}~\bibnamefont {Levy}}, \bibinfo {author}
  {\bibfnamefont {N.~A.}\ \bibnamefont {Mortensen}},\ and\ \bibinfo {author}
  {\bibfnamefont {A.}~\bibnamefont {Kristensen}},\ }\bibfield  {title}
  {\bibinfo {title} {{Resonant laser printing of structural colors on
  high-index dielectric metasurfaces}},\ }\href
  {https://doi.org/10.1126/sciadv.1602487} {\bibfield  {journal} {\bibinfo
  {journal} {Science Advances}\ }\textbf {\bibinfo {volume} {3}},\ \bibinfo
  {pages} {1} (\bibinfo {year} {2017})}\BibitemShut {NoStop}%
\bibitem [{\citenamefont {Hentschel}\ \emph {et~al.}(2017)\citenamefont
  {Hentschel}, \citenamefont {Sch{\"{a}}ferling}, \citenamefont {Duan},
  \citenamefont {Giessen},\ and\ \citenamefont {Liu}}]{Hentschel2017}%
  \BibitemOpen
  \bibfield  {author} {\bibinfo {author} {\bibfnamefont {M.}~\bibnamefont
  {Hentschel}}, \bibinfo {author} {\bibfnamefont {M.}~\bibnamefont
  {Sch{\"{a}}ferling}}, \bibinfo {author} {\bibfnamefont {X.}~\bibnamefont
  {Duan}}, \bibinfo {author} {\bibfnamefont {H.}~\bibnamefont {Giessen}},\ and\
  \bibinfo {author} {\bibfnamefont {N.}~\bibnamefont {Liu}},\ }\bibfield
  {title} {\bibinfo {title} {{Chiral plasmonics}},\ }\href
  {https://doi.org/10.1126/sciadv.1602735} {\bibfield  {journal} {\bibinfo
  {journal} {Science Advances}\ }\textbf {\bibinfo {volume} {3}},\ \bibinfo
  {pages} {e1602735} (\bibinfo {year} {2017})}\BibitemShut {NoStop}%
\bibitem [{\citenamefont {Aieta}\ \emph {et~al.}(2015)\citenamefont {Aieta},
  \citenamefont {Kats}, \citenamefont {Genevet},\ and\ \citenamefont
  {Capasso}}]{Aieta2015a}%
  \BibitemOpen
  \bibfield  {author} {\bibinfo {author} {\bibfnamefont {F.}~\bibnamefont
  {Aieta}}, \bibinfo {author} {\bibfnamefont {M.~A.}\ \bibnamefont {Kats}},
  \bibinfo {author} {\bibfnamefont {P.}~\bibnamefont {Genevet}},\ and\ \bibinfo
  {author} {\bibfnamefont {F.}~\bibnamefont {Capasso}},\ }\bibfield  {title}
  {\bibinfo {title} {{Multiwavelength achromatic metasurfaces by dispersive
  phase compensation}},\ }\href {https://doi.org/10.1126/science.aaa2494}
  {\bibfield  {journal} {\bibinfo  {journal} {Science}\ }\textbf {\bibinfo
  {volume} {347}},\ \bibinfo {pages} {1342} (\bibinfo {year} {2015})},\ \Eprint
  {https://arxiv.org/abs/1411.3966} {arXiv:1411.3966} \BibitemShut {NoStop}%
\bibitem [{\citenamefont {Lin}\ \emph {et~al.}(2014)\citenamefont {Lin},
  \citenamefont {Fan}, \citenamefont {Hasman},\ and\ \citenamefont
  {Brongersma}}]{Lin2014}%
  \BibitemOpen
  \bibfield  {author} {\bibinfo {author} {\bibfnamefont {D.}~\bibnamefont
  {Lin}}, \bibinfo {author} {\bibfnamefont {P.}~\bibnamefont {Fan}}, \bibinfo
  {author} {\bibfnamefont {E.}~\bibnamefont {Hasman}},\ and\ \bibinfo {author}
  {\bibfnamefont {M.~L.}\ \bibnamefont {Brongersma}},\ }\bibfield  {title}
  {\bibinfo {title} {{Dielectric gradient metasurface optical elements}},\
  }\href {https://doi.org/10.1126/science.1253213} {\bibfield  {journal}
  {\bibinfo  {journal} {Science}\ }\textbf {\bibinfo {volume} {345}},\ \bibinfo
  {pages} {298} (\bibinfo {year} {2014})}\BibitemShut {NoStop}%
\bibitem [{\citenamefont {Genevet}\ \emph {et~al.}(2017)\citenamefont
  {Genevet}, \citenamefont {Capasso}, \citenamefont {Aieta}, \citenamefont
  {Khorasaninejad},\ and\ \citenamefont {Devlin}}]{Genevet2017}%
  \BibitemOpen
  \bibfield  {author} {\bibinfo {author} {\bibfnamefont {P.}~\bibnamefont
  {Genevet}}, \bibinfo {author} {\bibfnamefont {F.}~\bibnamefont {Capasso}},
  \bibinfo {author} {\bibfnamefont {F.}~\bibnamefont {Aieta}}, \bibinfo
  {author} {\bibfnamefont {M.}~\bibnamefont {Khorasaninejad}},\ and\ \bibinfo
  {author} {\bibfnamefont {R.}~\bibnamefont {Devlin}},\ }\bibfield  {title}
  {\bibinfo {title} {{Recent advances in planar optics: from plasmonic to
  dielectric metasurfaces}},\ }\href {https://doi.org/10.1364/OPTICA.4.000139}
  {\bibfield  {journal} {\bibinfo  {journal} {Optica}\ }\textbf {\bibinfo
  {volume} {4}},\ \bibinfo {pages} {139} (\bibinfo {year} {2017})}\BibitemShut
  {NoStop}%
\bibitem [{\citenamefont {Tretyakov}(2017)}]{Tretyakov}%
  \BibitemOpen
  \bibfield  {author} {\bibinfo {author} {\bibfnamefont {S.~A.}\ \bibnamefont
  {Tretyakov}},\ }\bibfield  {title} {\bibinfo {title} {{A personal view on the
  origins and developments of the metamaterial concept}},\ }\href
  {https://doi.org/10.1088/2040-8986/19/1/013002} {\bibfield  {journal}
  {\bibinfo  {journal} {Journal of Optics}\ }\textbf {\bibinfo {volume} {19}},\
  \bibinfo {pages} {013002} (\bibinfo {year} {2017})}\BibitemShut {NoStop}%
\bibitem [{\citenamefont {Staude}\ \emph {et~al.}(2019)\citenamefont {Staude},
  \citenamefont {Pertsch},\ and\ \citenamefont {Kivshar}}]{Staude2019}%
  \BibitemOpen
  \bibfield  {author} {\bibinfo {author} {\bibfnamefont {I.}~\bibnamefont
  {Staude}}, \bibinfo {author} {\bibfnamefont {T.}~\bibnamefont {Pertsch}},\
  and\ \bibinfo {author} {\bibfnamefont {Y.~S.}\ \bibnamefont {Kivshar}},\
  }\bibfield  {title} {\bibinfo {title} {{All-Dielectric Resonant Meta-Optics
  Lightens up}},\ }\bibfield  {journal} {\bibinfo  {journal} {ACS Photonics}\
  }\href {https://doi.org/10.1021/acsphotonics.8b01326}
  {10.1021/acsphotonics.8b01326} (\bibinfo {year} {2019})\BibitemShut {NoStop}%
\bibitem [{\citenamefont {Kenanakis}\ \emph {et~al.}(2015)\citenamefont
  {Kenanakis}, \citenamefont {Xomalis}, \citenamefont {Selimis}, \citenamefont
  {Vamvakaki}, \citenamefont {Farsari}, \citenamefont {Kafesaki}, \citenamefont
  {Soukoulis},\ and\ \citenamefont {Economou}}]{Kenanakis2015}%
  \BibitemOpen
  \bibfield  {author} {\bibinfo {author} {\bibfnamefont {G.}~\bibnamefont
  {Kenanakis}}, \bibinfo {author} {\bibfnamefont {A.}~\bibnamefont {Xomalis}},
  \bibinfo {author} {\bibfnamefont {A.}~\bibnamefont {Selimis}}, \bibinfo
  {author} {\bibfnamefont {M.}~\bibnamefont {Vamvakaki}}, \bibinfo {author}
  {\bibfnamefont {M.}~\bibnamefont {Farsari}}, \bibinfo {author} {\bibfnamefont
  {M.}~\bibnamefont {Kafesaki}}, \bibinfo {author} {\bibfnamefont {C.~M.}\
  \bibnamefont {Soukoulis}},\ and\ \bibinfo {author} {\bibfnamefont {E.~N.}\
  \bibnamefont {Economou}},\ }\bibfield  {title} {\bibinfo {title}
  {{Three-Dimensional Infrared Metamaterial with Asymmetric Transmission}},\
  }\href {https://doi.org/10.1021/ph5003818} {\bibfield  {journal} {\bibinfo
  {journal} {ACS Photonics}\ }\textbf {\bibinfo {volume} {2}},\ \bibinfo
  {pages} {287} (\bibinfo {year} {2015})}\BibitemShut {NoStop}%
\bibitem [{\citenamefont {Liu}\ and\ \citenamefont {Lalanne}(2010)}]{Liu2010a}%
  \BibitemOpen
  \bibfield  {author} {\bibinfo {author} {\bibfnamefont {H.}~\bibnamefont
  {Liu}}\ and\ \bibinfo {author} {\bibfnamefont {P.}~\bibnamefont {Lalanne}},\
  }\bibfield  {title} {\bibinfo {title} {{Comprehensive microscopic model of
  the extraordinary optical transmission}},\ }\href
  {https://doi.org/10.1364/JOSAA.27.002542} {\bibfield  {journal} {\bibinfo
  {journal} {Journal of the Optical Society of America A}\ }\textbf {\bibinfo
  {volume} {27}},\ \bibinfo {pages} {2542} (\bibinfo {year}
  {2010})}\BibitemShut {NoStop}%
\bibitem [{\citenamefont {Poddubny}\ \emph {et~al.}(2016)\citenamefont
  {Poddubny}, \citenamefont {Iorsh},\ and\ \citenamefont
  {Sukhorukov}}]{Poddubny2016}%
  \BibitemOpen
  \bibfield  {author} {\bibinfo {author} {\bibfnamefont {A.~N.}\ \bibnamefont
  {Poddubny}}, \bibinfo {author} {\bibfnamefont {I.~V.}\ \bibnamefont
  {Iorsh}},\ and\ \bibinfo {author} {\bibfnamefont {A.~A.}\ \bibnamefont
  {Sukhorukov}},\ }\bibfield  {title} {\bibinfo {title} {{Generation of
  Photon-Plasmon Quantum States in Nonlinear Hyperbolic Metamaterials}},\
  }\href {https://doi.org/10.1103/PhysRevLett.117.123901} {\bibfield  {journal}
  {\bibinfo  {journal} {Physical Review Letters}\ }\textbf {\bibinfo {volume}
  {117}},\ \bibinfo {pages} {123901} (\bibinfo {year} {2016})}\BibitemShut
  {NoStop}%
\bibitem [{\citenamefont {Limonov}\ \emph {et~al.}(2017)\citenamefont
  {Limonov}, \citenamefont {Rybin}, \citenamefont {Poddubny},\ and\
  \citenamefont {Kivshar}}]{Limonov2017}%
  \BibitemOpen
  \bibfield  {author} {\bibinfo {author} {\bibfnamefont {M.~F.}\ \bibnamefont
  {Limonov}}, \bibinfo {author} {\bibfnamefont {M.~V.}\ \bibnamefont {Rybin}},
  \bibinfo {author} {\bibfnamefont {A.~N.}\ \bibnamefont {Poddubny}},\ and\
  \bibinfo {author} {\bibfnamefont {Y.~S.}\ \bibnamefont {Kivshar}},\
  }\bibfield  {title} {\bibinfo {title} {{Fano resonances in photonics}},\
  }\href {https://doi.org/10.1038/NPHOTON.2017.142} {\bibfield  {journal}
  {\bibinfo  {journal} {Nature Photonics}\ }\textbf {\bibinfo {volume} {11}},\
  \bibinfo {pages} {543} (\bibinfo {year} {2017})}\BibitemShut {NoStop}%
\bibitem [{\citenamefont {Wang}\ \emph {et~al.}(2018)\citenamefont {Wang},
  \citenamefont {Titchener}, \citenamefont {Kruk}, \citenamefont {Xu},
  \citenamefont {Chung}, \citenamefont {Parry}, \citenamefont {Kravchenko},
  \citenamefont {Chen}, \citenamefont {Solntsev}, \citenamefont {Kivshar},
  \citenamefont {Neshev},\ and\ \citenamefont {Sukhorukov}}]{Wang2018}%
  \BibitemOpen
  \bibfield  {author} {\bibinfo {author} {\bibfnamefont {K.}~\bibnamefont
  {Wang}}, \bibinfo {author} {\bibfnamefont {J.~G.}\ \bibnamefont {Titchener}},
  \bibinfo {author} {\bibfnamefont {S.~S.}\ \bibnamefont {Kruk}}, \bibinfo
  {author} {\bibfnamefont {L.}~\bibnamefont {Xu}}, \bibinfo {author}
  {\bibfnamefont {H.-p.}\ \bibnamefont {Chung}}, \bibinfo {author}
  {\bibfnamefont {M.}~\bibnamefont {Parry}}, \bibinfo {author} {\bibfnamefont
  {I.~I.}\ \bibnamefont {Kravchenko}}, \bibinfo {author} {\bibfnamefont
  {Y.-h.}\ \bibnamefont {Chen}}, \bibinfo {author} {\bibfnamefont {A.~S.}\
  \bibnamefont {Solntsev}}, \bibinfo {author} {\bibfnamefont {Y.~S.}\
  \bibnamefont {Kivshar}}, \bibinfo {author} {\bibfnamefont {D.~N.}\
  \bibnamefont {Neshev}},\ and\ \bibinfo {author} {\bibfnamefont {A.~A.}\
  \bibnamefont {Sukhorukov}},\ }\bibfield  {title} {\bibinfo {title} {{Quantum
  metasurface for multiphoton interference and state reconstruction}},\ }\href
  {https://doi.org/10.1126/science.aat8196} {\bibfield  {journal} {\bibinfo
  {journal} {Science}\ }\textbf {\bibinfo {volume} {361}},\ \bibinfo {pages}
  {1104} (\bibinfo {year} {2018})}\BibitemShut {NoStop}%
\bibitem [{\citenamefont {Wang}\ \emph {et~al.}(2019)\citenamefont {Wang},
  \citenamefont {Suchkov}, \citenamefont {Titchener}, \citenamefont {Szameit},\
  and\ \citenamefont {Sukhorukov}}]{Wang2019}%
  \BibitemOpen
  \bibfield  {author} {\bibinfo {author} {\bibfnamefont {K.}~\bibnamefont
  {Wang}}, \bibinfo {author} {\bibfnamefont {S.~V.}\ \bibnamefont {Suchkov}},
  \bibinfo {author} {\bibfnamefont {J.~G.}\ \bibnamefont {Titchener}}, \bibinfo
  {author} {\bibfnamefont {A.}~\bibnamefont {Szameit}},\ and\ \bibinfo {author}
  {\bibfnamefont {A.~A.}\ \bibnamefont {Sukhorukov}},\ }\bibfield  {title}
  {\bibinfo {title} {{Inline detection and reconstruction of multiphoton
  quantum states}},\ }\href {https://doi.org/10.1364/OPTICA.6.000041}
  {\bibfield  {journal} {\bibinfo  {journal} {Optica}\ }\textbf {\bibinfo
  {volume} {6}},\ \bibinfo {pages} {41} (\bibinfo {year} {2019})}\BibitemShut
  {NoStop}%
\bibitem [{\citenamefont {Yesilkoy}\ \emph {et~al.}(2019)\citenamefont
  {Yesilkoy}, \citenamefont {Arvelo}, \citenamefont {Jahani}, \citenamefont
  {Liu}, \citenamefont {Tittl}, \citenamefont {Cevher}, \citenamefont
  {Kivshar},\ and\ \citenamefont {Altug}}]{Yesilkoy2019}%
  \BibitemOpen
  \bibfield  {author} {\bibinfo {author} {\bibfnamefont {F.}~\bibnamefont
  {Yesilkoy}}, \bibinfo {author} {\bibfnamefont {E.~R.}\ \bibnamefont
  {Arvelo}}, \bibinfo {author} {\bibfnamefont {Y.}~\bibnamefont {Jahani}},
  \bibinfo {author} {\bibfnamefont {M.}~\bibnamefont {Liu}}, \bibinfo {author}
  {\bibfnamefont {A.}~\bibnamefont {Tittl}}, \bibinfo {author} {\bibfnamefont
  {V.}~\bibnamefont {Cevher}}, \bibinfo {author} {\bibfnamefont
  {Y.}~\bibnamefont {Kivshar}},\ and\ \bibinfo {author} {\bibfnamefont
  {H.}~\bibnamefont {Altug}},\ }\bibfield  {title} {\bibinfo {title}
  {{Ultrasensitive hyperspectral imaging and biodetection enabled by dielectric
  metasurfaces}},\ }\href {https://doi.org/10.1038/s41566-019-0394-6}
  {\bibfield  {journal} {\bibinfo  {journal} {Nature Photonics}\ }\textbf
  {\bibinfo {volume} {13}},\ \bibinfo {pages} {390} (\bibinfo {year}
  {2019})}\BibitemShut {NoStop}%
\bibitem [{\citenamefont {Do}\ \emph {et~al.}(2013)\citenamefont {Do},
  \citenamefont {Park}, \citenamefont {Hwang}, \citenamefont {Lee},
  \citenamefont {Ju},\ and\ \citenamefont {Choi}}]{Do2013}%
  \BibitemOpen
  \bibfield  {author} {\bibinfo {author} {\bibfnamefont {Y.~S.}\ \bibnamefont
  {Do}}, \bibinfo {author} {\bibfnamefont {J.~H.}\ \bibnamefont {Park}},
  \bibinfo {author} {\bibfnamefont {B.~Y.}\ \bibnamefont {Hwang}}, \bibinfo
  {author} {\bibfnamefont {S.-M.}\ \bibnamefont {Lee}}, \bibinfo {author}
  {\bibfnamefont {B.-K.}\ \bibnamefont {Ju}},\ and\ \bibinfo {author}
  {\bibfnamefont {K.~C.}\ \bibnamefont {Choi}},\ }\bibfield  {title} {\bibinfo
  {title} {{Color Filters: Plasmonic Color Filter and its Fabrication for
  Large-Area Applications}},\ }\href {https://doi.org/10.1002/adom.201370010}
  {\bibfield  {journal} {\bibinfo  {journal} {Advanced Optical Materials}\
  }\textbf {\bibinfo {volume} {1}},\ \bibinfo {pages} {109} (\bibinfo {year}
  {2013})}\BibitemShut {NoStop}%
\bibitem [{\citenamefont {Forouzmand}\ and\ \citenamefont
  {Mosallaei}(2018)}]{Forouzmand2018}%
  \BibitemOpen
  \bibfield  {author} {\bibinfo {author} {\bibfnamefont {A.}~\bibnamefont
  {Forouzmand}}\ and\ \bibinfo {author} {\bibfnamefont {H.}~\bibnamefont
  {Mosallaei}},\ }\bibfield  {title} {\bibinfo {title} {{Composite Multilayer
  Shared-Aperture Nanostructures: A Functional Multispectral Control}},\ }\href
  {https://doi.org/10.1021/acsphotonics.7b01441} {\bibfield  {journal}
  {\bibinfo  {journal} {ACS Photonics}\ }\textbf {\bibinfo {volume} {5}},\
  \bibinfo {pages} {1427} (\bibinfo {year} {2018})}\BibitemShut {NoStop}%
\bibitem [{\citenamefont {Jin}\ \emph {et~al.}(2018)\citenamefont {Jin},
  \citenamefont {Dong}, \citenamefont {Mei}, \citenamefont {Yu}, \citenamefont
  {Wei}, \citenamefont {Pan}, \citenamefont {Rezaei}, \citenamefont {Li},
  \citenamefont {Kuznetsov}, \citenamefont {Kivshar}, \citenamefont {Yang},\
  and\ \citenamefont {Qiu}}]{Jin2018}%
  \BibitemOpen
  \bibfield  {author} {\bibinfo {author} {\bibfnamefont {L.}~\bibnamefont
  {Jin}}, \bibinfo {author} {\bibfnamefont {Z.}~\bibnamefont {Dong}}, \bibinfo
  {author} {\bibfnamefont {S.}~\bibnamefont {Mei}}, \bibinfo {author}
  {\bibfnamefont {Y.~F.}\ \bibnamefont {Yu}}, \bibinfo {author} {\bibfnamefont
  {Z.}~\bibnamefont {Wei}}, \bibinfo {author} {\bibfnamefont {Z.}~\bibnamefont
  {Pan}}, \bibinfo {author} {\bibfnamefont {S.~D.}\ \bibnamefont {Rezaei}},
  \bibinfo {author} {\bibfnamefont {X.}~\bibnamefont {Li}}, \bibinfo {author}
  {\bibfnamefont {A.~I.}\ \bibnamefont {Kuznetsov}}, \bibinfo {author}
  {\bibfnamefont {Y.~S.}\ \bibnamefont {Kivshar}}, \bibinfo {author}
  {\bibfnamefont {J.~K.~W.}\ \bibnamefont {Yang}},\ and\ \bibinfo {author}
  {\bibfnamefont {C.-w.}\ \bibnamefont {Qiu}},\ }\bibfield  {title} {\bibinfo
  {title} {{Noninterleaved Metasurface for (2 6 -1) Spin- and
  Wavelength-Encoded Holograms}},\ }\href
  {https://doi.org/10.1021/acs.nanolett.8b04246} {\bibfield  {journal}
  {\bibinfo  {journal} {Nano Letters}\ }\textbf {\bibinfo {volume} {18}},\
  \bibinfo {pages} {8016} (\bibinfo {year} {2018})}\BibitemShut {NoStop}%
\bibitem [{\citenamefont {Li}\ \emph {et~al.}(2016)\citenamefont {Li},
  \citenamefont {Chen}, \citenamefont {Li}, \citenamefont {Zhang},
  \citenamefont {Pu}, \citenamefont {Zhao}, \citenamefont {Ma}, \citenamefont
  {Wang}, \citenamefont {Hong},\ and\ \citenamefont {Luo}}]{Li2016a}%
  \BibitemOpen
  \bibfield  {author} {\bibinfo {author} {\bibfnamefont {X.}~\bibnamefont
  {Li}}, \bibinfo {author} {\bibfnamefont {L.}~\bibnamefont {Chen}}, \bibinfo
  {author} {\bibfnamefont {Y.}~\bibnamefont {Li}}, \bibinfo {author}
  {\bibfnamefont {X.}~\bibnamefont {Zhang}}, \bibinfo {author} {\bibfnamefont
  {M.}~\bibnamefont {Pu}}, \bibinfo {author} {\bibfnamefont {Z.}~\bibnamefont
  {Zhao}}, \bibinfo {author} {\bibfnamefont {X.}~\bibnamefont {Ma}}, \bibinfo
  {author} {\bibfnamefont {Y.}~\bibnamefont {Wang}}, \bibinfo {author}
  {\bibfnamefont {M.}~\bibnamefont {Hong}},\ and\ \bibinfo {author}
  {\bibfnamefont {X.}~\bibnamefont {Luo}},\ }\bibfield  {title} {\bibinfo
  {title} {{Multicolor 3D meta-holography by broadband plasmonic modulation}},\
  }\href {https://doi.org/10.1126/sciadv.1601102} {\bibfield  {journal}
  {\bibinfo  {journal} {Science Advances}\ }\textbf {\bibinfo {volume} {2}},\
  \bibinfo {pages} {e1601102} (\bibinfo {year} {2016})}\BibitemShut {NoStop}%
\bibitem [{\citenamefont {Zhou}\ \emph {et~al.}(2018)\citenamefont {Zhou},
  \citenamefont {Kravchenko}, \citenamefont {Wang}, \citenamefont {Nolen},
  \citenamefont {Gu},\ and\ \citenamefont {Valentine}}]{Zhou2018a}%
  \BibitemOpen
  \bibfield  {author} {\bibinfo {author} {\bibfnamefont {Y.}~\bibnamefont
  {Zhou}}, \bibinfo {author} {\bibfnamefont {I.~I.}\ \bibnamefont
  {Kravchenko}}, \bibinfo {author} {\bibfnamefont {H.}~\bibnamefont {Wang}},
  \bibinfo {author} {\bibfnamefont {J.~R.}\ \bibnamefont {Nolen}}, \bibinfo
  {author} {\bibfnamefont {G.}~\bibnamefont {Gu}},\ and\ \bibinfo {author}
  {\bibfnamefont {J.}~\bibnamefont {Valentine}},\ }\bibfield  {title} {\bibinfo
  {title} {{Multilayer Noninteracting Dielectric Metasurfaces for
  Multiwavelength Metaoptics}},\ }\href
  {https://doi.org/10.1021/acs.nanolett.8b03017} {\bibfield  {journal}
  {\bibinfo  {journal} {Nano Letters}\ }\textbf {\bibinfo {volume} {18}},\
  \bibinfo {pages} {7529} (\bibinfo {year} {2018})}\BibitemShut {NoStop}%
\bibitem [{\citenamefont {Arbabi}\ \emph {et~al.}(2016)\citenamefont {Arbabi},
  \citenamefont {Arbabi}, \citenamefont {Kamali}, \citenamefont {Horie},
  \citenamefont {Han},\ and\ \citenamefont {Faraon}}]{Arbabi2016a}%
  \BibitemOpen
  \bibfield  {author} {\bibinfo {author} {\bibfnamefont {A.}~\bibnamefont
  {Arbabi}}, \bibinfo {author} {\bibfnamefont {E.}~\bibnamefont {Arbabi}},
  \bibinfo {author} {\bibfnamefont {S.~M.}\ \bibnamefont {Kamali}}, \bibinfo
  {author} {\bibfnamefont {Y.}~\bibnamefont {Horie}}, \bibinfo {author}
  {\bibfnamefont {S.}~\bibnamefont {Han}},\ and\ \bibinfo {author}
  {\bibfnamefont {A.}~\bibnamefont {Faraon}},\ }\bibfield  {title} {\bibinfo
  {title} {{Miniature optical planar camera based on a wide-angle metasurface
  doublet corrected for monochromatic aberrations}},\ }\href
  {https://doi.org/10.1038/ncomms13682} {\bibfield  {journal} {\bibinfo
  {journal} {Nature Communications}\ }\textbf {\bibinfo {volume} {7}},\
  \bibinfo {pages} {13682} (\bibinfo {year} {2016})},\ \Eprint
  {https://arxiv.org/abs/arXiv:1604.06160v2} {arXiv:arXiv:1604.06160v2}
  \BibitemShut {NoStop}%
\bibitem [{\citenamefont {Kuznetsov}\ \emph {et~al.}(2015)\citenamefont
  {Kuznetsov}, \citenamefont {Astafev}, \citenamefont {Beruete},\ and\
  \citenamefont {Navarro-C{\'{i}}a}}]{Kuznetsov2015}%
  \BibitemOpen
  \bibfield  {author} {\bibinfo {author} {\bibfnamefont {S.~a.}\ \bibnamefont
  {Kuznetsov}}, \bibinfo {author} {\bibfnamefont {M.~a.}\ \bibnamefont
  {Astafev}}, \bibinfo {author} {\bibfnamefont {M.}~\bibnamefont {Beruete}},\
  and\ \bibinfo {author} {\bibfnamefont {M.}~\bibnamefont
  {Navarro-C{\'{i}}a}},\ }\bibfield  {title} {\bibinfo {title} {{Planar
  Holographic Metasurfaces for Terahertz Focusing}},\ }\href
  {https://doi.org/10.1038/srep07738} {\bibfield  {journal} {\bibinfo
  {journal} {Scientific Reports}\ }\textbf {\bibinfo {volume} {5}},\ \bibinfo
  {pages} {7738} (\bibinfo {year} {2015})}\BibitemShut {NoStop}%
\bibitem [{\citenamefont {Hsu}\ \emph {et~al.}(2013)\citenamefont {Hsu},
  \citenamefont {Zhen}, \citenamefont {Lee}, \citenamefont {Chua},
  \citenamefont {Johnson}, \citenamefont {Joannopoulos},\ and\ \citenamefont
  {Solja{\v{c}}i{\'{c}}}}]{Hsu2013}%
  \BibitemOpen
  \bibfield  {author} {\bibinfo {author} {\bibfnamefont {C.~W.}\ \bibnamefont
  {Hsu}}, \bibinfo {author} {\bibfnamefont {B.}~\bibnamefont {Zhen}}, \bibinfo
  {author} {\bibfnamefont {J.}~\bibnamefont {Lee}}, \bibinfo {author}
  {\bibfnamefont {S.-l.}\ \bibnamefont {Chua}}, \bibinfo {author}
  {\bibfnamefont {S.~G.}\ \bibnamefont {Johnson}}, \bibinfo {author}
  {\bibfnamefont {J.~D.}\ \bibnamefont {Joannopoulos}},\ and\ \bibinfo {author}
  {\bibfnamefont {M.}~\bibnamefont {Solja{\v{c}}i{\'{c}}}},\ }\bibfield
  {title} {\bibinfo {title} {{Observation of trapped light within the radiation
  continuum}},\ }\href {https://doi.org/10.1038/nature12289} {\bibfield
  {journal} {\bibinfo  {journal} {Nature}\ }\textbf {\bibinfo {volume} {499}},\
  \bibinfo {pages} {188} (\bibinfo {year} {2013})}\BibitemShut {NoStop}%
\bibitem [{\citenamefont {Monticone}\ and\ \citenamefont
  {Al{\`{u}}}(2014)}]{Monticone2014}%
  \BibitemOpen
  \bibfield  {author} {\bibinfo {author} {\bibfnamefont {F.}~\bibnamefont
  {Monticone}}\ and\ \bibinfo {author} {\bibfnamefont {A.}~\bibnamefont
  {Al{\`{u}}}},\ }\bibfield  {title} {\bibinfo {title} {{Embedded Photonic
  Eigenvalues in 3D Nanostructures}},\ }\href
  {https://doi.org/10.1103/PhysRevLett.112.213903} {\bibfield  {journal}
  {\bibinfo  {journal} {Physical Review Letters}\ }\textbf {\bibinfo {volume}
  {112}},\ \bibinfo {pages} {213903} (\bibinfo {year} {2014})}\BibitemShut
  {NoStop}%
\bibitem [{\citenamefont {Hsu}\ \emph {et~al.}(2016)\citenamefont {Hsu},
  \citenamefont {Zhen}, \citenamefont {Stone}, \citenamefont {Joannopoulos},\
  and\ \citenamefont {Solja{\v{c}}i{\'{c}}}}]{Hsu2016}%
  \BibitemOpen
  \bibfield  {author} {\bibinfo {author} {\bibfnamefont {C.~W.}\ \bibnamefont
  {Hsu}}, \bibinfo {author} {\bibfnamefont {B.}~\bibnamefont {Zhen}}, \bibinfo
  {author} {\bibfnamefont {A.~D.}\ \bibnamefont {Stone}}, \bibinfo {author}
  {\bibfnamefont {J.~D.}\ \bibnamefont {Joannopoulos}},\ and\ \bibinfo {author}
  {\bibfnamefont {M.}~\bibnamefont {Solja{\v{c}}i{\'{c}}}},\ }\bibfield
  {title} {\bibinfo {title} {{Bound states in the continuum}},\ }\href
  {https://doi.org/10.1038/natrevmats.2016.48
  http://10.0.4.14/natrevmats.2016.48} {\bibfield  {journal} {\bibinfo
  {journal} {Nature Reviews Materials}\ }\textbf {\bibinfo {volume} {1}},\
  \bibinfo {pages} {16048} (\bibinfo {year} {2016})}\BibitemShut {NoStop}%
\bibitem [{\citenamefont {Cerjan}\ \emph {et~al.}(2019)\citenamefont {Cerjan},
  \citenamefont {Hsu},\ and\ \citenamefont {Rechtsman}}]{Cerjan2019}%
  \BibitemOpen
  \bibfield  {author} {\bibinfo {author} {\bibfnamefont {A.}~\bibnamefont
  {Cerjan}}, \bibinfo {author} {\bibfnamefont {C.~W.}\ \bibnamefont {Hsu}},\
  and\ \bibinfo {author} {\bibfnamefont {M.~C.}\ \bibnamefont {Rechtsman}},\
  }\bibfield  {title} {\bibinfo {title} {{Bound States in the Continuum through
  Environmental Design}},\ }\href
  {https://doi.org/10.1103/PhysRevLett.123.023902} {\bibfield  {journal}
  {\bibinfo  {journal} {Physical Review Letters}\ }\textbf {\bibinfo {volume}
  {123}},\ \bibinfo {pages} {023902} (\bibinfo {year} {2019})}\BibitemShut
  {NoStop}%
\bibitem [{\citenamefont {V{\'{a}}zquez-Guardado}\ and\ \citenamefont
  {Chanda}(2018)}]{Vazquez-guardado2018}%
  \BibitemOpen
  \bibfield  {author} {\bibinfo {author} {\bibfnamefont {A.}~\bibnamefont
  {V{\'{a}}zquez-Guardado}}\ and\ \bibinfo {author} {\bibfnamefont
  {D.}~\bibnamefont {Chanda}},\ }\bibfield  {title} {\bibinfo {title}
  {{Superchiral Light Generation on Degenerate Achiral Surfaces}},\ }\href
  {https://doi.org/10.1103/PhysRevLett.120.137601} {\bibfield  {journal}
  {\bibinfo  {journal} {Physical Review Letters}\ }\textbf {\bibinfo {volume}
  {120}},\ \bibinfo {pages} {137601} (\bibinfo {year} {2018})}\BibitemShut
  {NoStop}%
\bibitem [{\citenamefont {Duggan}\ \emph {et~al.}(2019)\citenamefont {Duggan},
  \citenamefont {del Pino}, \citenamefont {Verhagen},\ and\ \citenamefont
  {Al{\`{u}}}}]{Duggan2019}%
  \BibitemOpen
  \bibfield  {author} {\bibinfo {author} {\bibfnamefont {R.}~\bibnamefont
  {Duggan}}, \bibinfo {author} {\bibfnamefont {J.}~\bibnamefont {del Pino}},
  \bibinfo {author} {\bibfnamefont {E.}~\bibnamefont {Verhagen}},\ and\
  \bibinfo {author} {\bibfnamefont {A.}~\bibnamefont {Al{\`{u}}}},\ }\bibfield
  {title} {\bibinfo {title} {{Optomechanically Induced Birefringence and
  Optomechanically Induced Faraday Effect}},\ }\href
  {https://doi.org/10.1103/PhysRevLett.123.023602} {\bibfield  {journal}
  {\bibinfo  {journal} {Physical Review Letters}\ }\textbf {\bibinfo {volume}
  {123}},\ \bibinfo {pages} {023602} (\bibinfo {year} {2019})},\ \Eprint
  {https://arxiv.org/abs/1904.05463} {arXiv:1904.05463} \BibitemShut {NoStop}%
\bibitem [{\citenamefont {Berkhout}\ and\ \citenamefont
  {Koenderink}(2020)}]{Berkhout2020}%
  \BibitemOpen
  \bibfield  {author} {\bibinfo {author} {\bibfnamefont {A.}~\bibnamefont
  {Berkhout}}\ and\ \bibinfo {author} {\bibfnamefont {A.~F.}\ \bibnamefont
  {Koenderink}},\ }\bibfield  {title} {\bibinfo {title} {{A simple
  transfer-matrix model for metasurface multilayer systems}},\ }\href
  {https://doi.org/10.1515/nanoph-2020-0212} {\bibfield  {journal} {\bibinfo
  {journal} {Nanophotonics}\ }\textbf {\bibinfo {volume} {0}},\ \bibinfo
  {pages} {20200212} (\bibinfo {year} {2020})}\BibitemShut {NoStop}%
\bibitem [{\citenamefont {Chen}\ \emph {et~al.}(2019)\citenamefont {Chen},
  \citenamefont {Zhang}, \citenamefont {Li}, \citenamefont {Cheng},\ and\
  \citenamefont {Tian}}]{Chen2019a}%
  \BibitemOpen
  \bibfield  {author} {\bibinfo {author} {\bibfnamefont {S.}~\bibnamefont
  {Chen}}, \bibinfo {author} {\bibfnamefont {Y.}~\bibnamefont {Zhang}},
  \bibinfo {author} {\bibfnamefont {Z.}~\bibnamefont {Li}}, \bibinfo {author}
  {\bibfnamefont {H.}~\bibnamefont {Cheng}},\ and\ \bibinfo {author}
  {\bibfnamefont {J.}~\bibnamefont {Tian}},\ }\bibfield  {title} {\bibinfo
  {title} {{Empowered Layer Effects and Prominent Properties in Few‐Layer
  Metasurfaces}},\ }\href {https://doi.org/10.1002/adom.201801477} {\bibfield
  {journal} {\bibinfo  {journal} {Advanced Optical Materials}\ }\textbf
  {\bibinfo {volume} {1801477}},\ \bibinfo {pages} {1801477} (\bibinfo {year}
  {2019})}\BibitemShut {NoStop}%
\bibitem [{\citenamefont {Menzel}\ \emph {et~al.}(2016)\citenamefont {Menzel},
  \citenamefont {Sperrhake},\ and\ \citenamefont {Pertsch}}]{Menzel2015}%
  \BibitemOpen
  \bibfield  {author} {\bibinfo {author} {\bibfnamefont {C.}~\bibnamefont
  {Menzel}}, \bibinfo {author} {\bibfnamefont {J.}~\bibnamefont {Sperrhake}},\
  and\ \bibinfo {author} {\bibfnamefont {T.}~\bibnamefont {Pertsch}},\
  }\bibfield  {title} {\bibinfo {title} {{Efficient treatment of stacked
  metasurfaces for optimizing and enhancing the range of accessible optical
  functionalities}},\ }\href {https://doi.org/10.1103/PhysRevA.93.063832}
  {\bibfield  {journal} {\bibinfo  {journal} {Physical Review A}\ }\textbf
  {\bibinfo {volume} {93}},\ \bibinfo {pages} {063832} (\bibinfo {year}
  {2016})},\ \Eprint {https://arxiv.org/abs/1511.09239} {arXiv:1511.09239}
  \BibitemShut {NoStop}%
\bibitem [{\citenamefont {Zhao}\ \emph {et~al.}(2012)\citenamefont {Zhao},
  \citenamefont {Belkin},\ and\ \citenamefont {Al{\`{u}}}}]{Zhao2012a}%
  \BibitemOpen
  \bibfield  {author} {\bibinfo {author} {\bibfnamefont {Y.}~\bibnamefont
  {Zhao}}, \bibinfo {author} {\bibfnamefont {M.}~\bibnamefont {Belkin}},\ and\
  \bibinfo {author} {\bibfnamefont {A.}~\bibnamefont {Al{\`{u}}}},\ }\bibfield
  {title} {\bibinfo {title} {{Twisted optical metamaterials for planarized
  ultrathin broadband circular polarizers}},\ }\href
  {https://doi.org/10.1038/ncomms1877} {\bibfield  {journal} {\bibinfo
  {journal} {Nature Communications}\ }\textbf {\bibinfo {volume} {3}},\
  \bibinfo {pages} {870} (\bibinfo {year} {2012})}\BibitemShut {NoStop}%
\bibitem [{\citenamefont {Lin}\ \emph {et~al.}(2019)\citenamefont {Lin},
  \citenamefont {Sturmberg}, \citenamefont {Lin}, \citenamefont {Yang},
  \citenamefont {Zheng}, \citenamefont {Chong}, \citenamefont {de~Sterke},\
  and\ \citenamefont {Jia}}]{Lin2019}%
  \BibitemOpen
  \bibfield  {author} {\bibinfo {author} {\bibfnamefont {H.}~\bibnamefont
  {Lin}}, \bibinfo {author} {\bibfnamefont {B.~C.~P.}\ \bibnamefont
  {Sturmberg}}, \bibinfo {author} {\bibfnamefont {K.-T.}\ \bibnamefont {Lin}},
  \bibinfo {author} {\bibfnamefont {Y.}~\bibnamefont {Yang}}, \bibinfo {author}
  {\bibfnamefont {X.}~\bibnamefont {Zheng}}, \bibinfo {author} {\bibfnamefont
  {T.~K.}\ \bibnamefont {Chong}}, \bibinfo {author} {\bibfnamefont {C.~M.}\
  \bibnamefont {de~Sterke}},\ and\ \bibinfo {author} {\bibfnamefont
  {B.}~\bibnamefont {Jia}},\ }\bibfield  {title} {\bibinfo {title} {{A
  90-nm-thick graphene metamaterial for strong and extremely broadband
  absorption of unpolarized light}},\ }\href
  {https://doi.org/10.1038/s41566-019-0389-3} {\bibfield  {journal} {\bibinfo
  {journal} {Nature Photonics}\ }\textbf {\bibinfo {volume} {13}},\ \bibinfo
  {pages} {270} (\bibinfo {year} {2019})}\BibitemShut {NoStop}%
\bibitem [{\citenamefont {Sperrhake}\ \emph {et~al.}(2019)\citenamefont
  {Sperrhake}, \citenamefont {Decker}, \citenamefont {Falkner}, \citenamefont
  {Fasold}, \citenamefont {Kaiser}, \citenamefont {Staude},\ and\ \citenamefont
  {Pertsch}}]{Sperrhake2019}%
  \BibitemOpen
  \bibfield  {author} {\bibinfo {author} {\bibfnamefont {J.}~\bibnamefont
  {Sperrhake}}, \bibinfo {author} {\bibfnamefont {M.}~\bibnamefont {Decker}},
  \bibinfo {author} {\bibfnamefont {M.}~\bibnamefont {Falkner}}, \bibinfo
  {author} {\bibfnamefont {S.}~\bibnamefont {Fasold}}, \bibinfo {author}
  {\bibfnamefont {T.}~\bibnamefont {Kaiser}}, \bibinfo {author} {\bibfnamefont
  {I.}~\bibnamefont {Staude}},\ and\ \bibinfo {author} {\bibfnamefont
  {T.}~\bibnamefont {Pertsch}},\ }\bibfield  {title} {\bibinfo {title}
  {{Analyzing the polarization response of a chiral metasurface stack by
  semi-analytic modeling}},\ }\href {https://doi.org/10.1364/OE.27.001236}
  {\bibfield  {journal} {\bibinfo  {journal} {Optics Express}\ }\textbf
  {\bibinfo {volume} {27}},\ \bibinfo {pages} {1236} (\bibinfo {year}
  {2019})}\BibitemShut {NoStop}%
\bibitem [{\citenamefont {Yun}\ \emph {et~al.}(2018)\citenamefont {Yun},
  \citenamefont {Sung}, \citenamefont {Kim},\ and\ \citenamefont
  {Lee}}]{Yun2018}%
  \BibitemOpen
  \bibfield  {author} {\bibinfo {author} {\bibfnamefont {J.-G.}\ \bibnamefont
  {Yun}}, \bibinfo {author} {\bibfnamefont {J.}~\bibnamefont {Sung}}, \bibinfo
  {author} {\bibfnamefont {S.-J.}\ \bibnamefont {Kim}},\ and\ \bibinfo {author}
  {\bibfnamefont {B.}~\bibnamefont {Lee}},\ }\bibfield  {title} {\bibinfo
  {title} {{Simultaneous control of polarization and amplitude over broad
  bandwidth using multi-layered anisotropic metasurfaces}},\ }\href
  {https://doi.org/10.1364/OE.26.029826} {\bibfield  {journal} {\bibinfo
  {journal} {Optics Express}\ }\textbf {\bibinfo {volume} {26}},\ \bibinfo
  {pages} {29826} (\bibinfo {year} {2018})}\BibitemShut {NoStop}%
\bibitem [{\citenamefont {Datta}(1995)}]{Datta1995}%
  \BibitemOpen
  \bibfield  {author} {\bibinfo {author} {\bibfnamefont {S.}~\bibnamefont
  {Datta}},\ }\href@noop {} {\emph {\bibinfo {title} {{Electronic transport in
  mesoscopic systems}}}},\ \bibinfo {edition} {1st}\ ed.\ (\bibinfo
  {publisher} {Cambridge University Press},\ \bibinfo {year} {1995})\ p.\
  \bibinfo {pages} {393}\BibitemShut {NoStop}%
\bibitem [{\citenamefont {Robinson}\ and\ \citenamefont
  {Jeffery}(1995)}]{Robinson1995}%
  \BibitemOpen
  \bibfield  {author} {\bibinfo {author} {\bibfnamefont {S.~J.}\ \bibnamefont
  {Robinson}}\ and\ \bibinfo {author} {\bibfnamefont {M.}~\bibnamefont
  {Jeffery}},\ }\bibfield  {title} {\bibinfo {title} {{Conductance fluctuations
  in mesoscopic three-dimensional multiply connected normal-wire networks}},\
  }\href {https://doi.org/10.1103/PhysRevB.51.16807} {\bibfield  {journal}
  {\bibinfo  {journal} {Physical Review B}\ }\textbf {\bibinfo {volume} {51}},\
  \bibinfo {pages} {16807} (\bibinfo {year} {1995})}\BibitemShut {NoStop}%
\bibitem [{\citenamefont {B{\"{u}}ttiker}(1986)}]{Bttiker1986}%
  \BibitemOpen
  \bibfield  {author} {\bibinfo {author} {\bibfnamefont {M.}~\bibnamefont
  {B{\"{u}}ttiker}},\ }\bibfield  {title} {\bibinfo {title} {{Four-Terminal
  Phase-Coherent Conductance}},\ }\href
  {https://doi.org/10.1103/PhysRevLett.57.1761} {\bibfield  {journal} {\bibinfo
   {journal} {Physical Review Letters}\ }\textbf {\bibinfo {volume} {57}},\
  \bibinfo {pages} {1761} (\bibinfo {year} {1986})}\BibitemShut {NoStop}%
\bibitem [{\citenamefont {Nazarov}(2015)}]{Nazarov2015}%
  \BibitemOpen
  \bibfield  {author} {\bibinfo {author} {\bibfnamefont {Y.~V.}\ \bibnamefont
  {Nazarov}},\ }\bibfield  {title} {\bibinfo {title} {{Block-determinant
  formalism for an action of a multi-terminal scatterer}},\ }\href
  {https://doi.org/10.1016/j.physe.2015.08.007} {\bibfield  {journal} {\bibinfo
   {journal} {Physica E: Low-dimensional Systems and Nanostructures}\ }\textbf
  {\bibinfo {volume} {74}},\ \bibinfo {pages} {561} (\bibinfo {year}
  {2015})}\BibitemShut {NoStop}%
\bibitem [{\citenamefont {Shavit}\ and\ \citenamefont
  {Oreg}(2019)}]{Shavit2019}%
  \BibitemOpen
  \bibfield  {author} {\bibinfo {author} {\bibfnamefont {G.}~\bibnamefont
  {Shavit}}\ and\ \bibinfo {author} {\bibfnamefont {Y.}~\bibnamefont {Oreg}},\
  }\bibfield  {title} {\bibinfo {title} {{Fractional Conductance in Strongly
  Interacting 1D Systems}},\ }\href
  {https://doi.org/10.1103/PhysRevLett.123.036803} {\bibfield  {journal}
  {\bibinfo  {journal} {Physical Review Letters}\ }\textbf {\bibinfo {volume}
  {123}},\ \bibinfo {pages} {036803} (\bibinfo {year} {2019})}\BibitemShut
  {NoStop}%
\bibitem [{\citenamefont {Texier}\ and\ \citenamefont
  {Montambaux}(2016)}]{Texier2016}%
  \BibitemOpen
  \bibfield  {author} {\bibinfo {author} {\bibfnamefont {C.}~\bibnamefont
  {Texier}}\ and\ \bibinfo {author} {\bibfnamefont {G.}~\bibnamefont
  {Montambaux}},\ }\bibfield  {title} {\bibinfo {title} {{Reprint of:
  Four-terminal resistances in mesoscopic networks of metallic wires: Weak
  localisation and correlations}},\ }\href
  {https://doi.org/10.1016/j.physe.2016.02.041} {\bibfield  {journal} {\bibinfo
   {journal} {Physica E: Low-dimensional Systems and Nanostructures}\ }\textbf
  {\bibinfo {volume} {82}},\ \bibinfo {pages} {272} (\bibinfo {year}
  {2016})}\BibitemShut {NoStop}%
\bibitem [{\citenamefont {Li}(1994)}]{Li1994a}%
  \BibitemOpen
  \bibfield  {author} {\bibinfo {author} {\bibfnamefont {L.}~\bibnamefont
  {Li}},\ }\bibfield  {title} {\bibinfo {title} {{Bremmer series, R-matrix
  propagation algorithm, and numerical modeling of diffraction gratings}},\
  }\href {https://doi.org/10.1364/JOSAA.11.002829} {\bibfield  {journal}
  {\bibinfo  {journal} {Journal of the Optical Society of America A}\ }\textbf
  {\bibinfo {volume} {11}},\ \bibinfo {pages} {2829} (\bibinfo {year}
  {1994})}\BibitemShut {NoStop}%
\bibitem [{\citenamefont {Li}(1996)}]{Li1996}%
  \BibitemOpen
  \bibfield  {author} {\bibinfo {author} {\bibfnamefont {L.}~\bibnamefont
  {Li}},\ }\bibfield  {title} {\bibinfo {title} {{Formulation and comparison of
  two recursive matrix algorithms for modeling layered diffraction gratings}},\
  }\href {https://doi.org/10.1364/JOSAA.13.001024} {\bibfield  {journal}
  {\bibinfo  {journal} {Journal of the Optical Society of America A}\ }\textbf
  {\bibinfo {volume} {13}},\ \bibinfo {pages} {1024 } (\bibinfo {year}
  {1996})}\BibitemShut {NoStop}%
\bibitem [{\citenamefont {Buttiker}(1988)}]{Buttiker1988}%
  \BibitemOpen
  \bibfield  {author} {\bibinfo {author} {\bibfnamefont {M.}~\bibnamefont
  {Buttiker}},\ }\bibfield  {title} {\bibinfo {title} {{Symmetry of electrical
  conduction}},\ }\href {https://doi.org/10.1147/rd.323.0317} {\bibfield
  {journal} {\bibinfo  {journal} {IBM Journal of Research and Development}\
  }\textbf {\bibinfo {volume} {32}},\ \bibinfo {pages} {317} (\bibinfo {year}
  {1988})}\BibitemShut {NoStop}%
\bibitem [{\citenamefont {B{\"{u}}ttiker}\ \emph {et~al.}(1985)\citenamefont
  {B{\"{u}}ttiker}, \citenamefont {Imry}, \citenamefont {Landauer},\ and\
  \citenamefont {Pinhas}}]{Bttiker1985}%
  \BibitemOpen
  \bibfield  {author} {\bibinfo {author} {\bibfnamefont {M.}~\bibnamefont
  {B{\"{u}}ttiker}}, \bibinfo {author} {\bibfnamefont {Y.}~\bibnamefont
  {Imry}}, \bibinfo {author} {\bibfnamefont {R.}~\bibnamefont {Landauer}},\
  and\ \bibinfo {author} {\bibfnamefont {S.}~\bibnamefont {Pinhas}},\
  }\bibfield  {title} {\bibinfo {title} {{Generalized many-channel conductance
  formula with application to small rings}},\ }\href
  {https://doi.org/10.1103/PhysRevB.31.6207} {\bibfield  {journal} {\bibinfo
  {journal} {Physical Review B}\ }\textbf {\bibinfo {volume} {31}},\ \bibinfo
  {pages} {6207} (\bibinfo {year} {1985})}\BibitemShut {NoStop}%
\bibitem [{\citenamefont {Simovski}(2007)}]{Simovski2007}%
  \BibitemOpen
  \bibfield  {author} {\bibinfo {author} {\bibfnamefont {C.~R.}\ \bibnamefont
  {Simovski}},\ }\bibfield  {title} {\bibinfo {title} {{Bloch material
  parameters of magneto-dielectric metamaterials and the concept of Bloch
  lattices}},\ }\href {https://doi.org/10.1016/j.metmat.2007.09.002} {\bibfield
   {journal} {\bibinfo  {journal} {Metamaterials}\ }\textbf {\bibinfo {volume}
  {1}},\ \bibinfo {pages} {62} (\bibinfo {year} {2007})}\BibitemShut {NoStop}%
\bibitem [{\citenamefont {Simovski}\ \emph {et~al.}(2007)\citenamefont
  {Simovski}, \citenamefont {Tretyakov},\ and\ \citenamefont
  {.}}]{Simovski2007a}%
  \BibitemOpen
  \bibfield  {author} {\bibinfo {author} {\bibfnamefont {C.~R.}\ \bibnamefont
  {Simovski}}, \bibinfo {author} {\bibfnamefont {S.~A.}\ \bibnamefont
  {Tretyakov}},\ and\ \bibinfo {author} {\bibnamefont {.}},\ }\bibfield
  {title} {\bibinfo {title} {{Local constitutive parameters of metamaterials
  from an effective-medium perspective}},\ }\href
  {https://doi.org/10.1103/PhysRevB.75.195111} {\bibfield  {journal} {\bibinfo
  {journal} {Physical Review B}\ }\textbf {\bibinfo {volume} {75}},\ \bibinfo
  {pages} {195111} (\bibinfo {year} {2007})}\BibitemShut {NoStop}%
\bibitem [{\citenamefont {Simovski}(2009)}]{Simovski2009}%
  \BibitemOpen
  \bibfield  {author} {\bibinfo {author} {\bibfnamefont {C.~R.}\ \bibnamefont
  {Simovski}},\ }\bibfield  {title} {\bibinfo {title} {{Material parameters of
  metamaterials (a Review)}},\ }\href
  {https://doi.org/10.1134/S0030400X09110101} {\bibfield  {journal} {\bibinfo
  {journal} {Optics and Spectroscopy}\ }\textbf {\bibinfo {volume} {107}},\
  \bibinfo {pages} {726} (\bibinfo {year} {2009})}\BibitemShut {NoStop}%
\bibitem [{\citenamefont {Menzel}\ \emph {et~al.}(2010)\citenamefont {Menzel},
  \citenamefont {Rockstuhl},\ and\ \citenamefont {Lederer}}]{Menzel2010}%
  \BibitemOpen
  \bibfield  {author} {\bibinfo {author} {\bibfnamefont {C.}~\bibnamefont
  {Menzel}}, \bibinfo {author} {\bibfnamefont {C.}~\bibnamefont {Rockstuhl}},\
  and\ \bibinfo {author} {\bibfnamefont {F.}~\bibnamefont {Lederer}},\
  }\bibfield  {title} {\bibinfo {title} {{Advanced Jones calculus for the
  classification of periodic metamaterials}},\ }\href
  {https://doi.org/10.1103/PhysRevA.82.053811} {\bibfield  {journal} {\bibinfo
  {journal} {Physical Review A - Atomic, Molecular, and Optical Physics}\
  }\textbf {\bibinfo {volume} {82}},\ \bibinfo {pages} {1} (\bibinfo {year}
  {2010})},\ \Eprint {https://arxiv.org/abs/1008.4117} {arXiv:1008.4117}
  \BibitemShut {NoStop}%
\bibitem [{\citenamefont {Redheffer}(1960)}]{Redheffer1960}%
  \BibitemOpen
  \bibfield  {author} {\bibinfo {author} {\bibfnamefont {R.~M.}\ \bibnamefont
  {Redheffer}},\ }\bibfield  {title} {\bibinfo {title} {{On a Certain Linear
  Fractional Transformation}},\ }\href {https://doi.org/10.1002/sapm1960391269}
  {\bibfield  {journal} {\bibinfo  {journal} {Journal of Mathematics and
  Physics}\ }\textbf {\bibinfo {volume} {39}},\ \bibinfo {pages} {269}
  (\bibinfo {year} {1960})}\BibitemShut {NoStop}%
\bibitem [{Note1()}]{Note1}%
  \BibitemOpen
  \bibinfo {note} {If it is invertible and its block-matrix elements do not
  take values $>1$. This is usually the case for physical systems including
  absorption.}\BibitemShut {Stop}%
\bibitem [{\citenamefont {Bremmer}(1951)}]{Bremmer1951}%
  \BibitemOpen
  \bibfield  {author} {\bibinfo {author} {\bibfnamefont {H.}~\bibnamefont
  {Bremmer}},\ }\bibfield  {title} {\bibinfo {title} {{The W.K.B. approximation
  as the first term of a geometric-optical series}},\ }\href
  {https://doi.org/10.1002/cpa.3160040111} {\bibfield  {journal} {\bibinfo
  {journal} {Communications on Pure and Applied Mathematics}\ }\textbf
  {\bibinfo {volume} {4}},\ \bibinfo {pages} {105} (\bibinfo {year}
  {1951})}\BibitemShut {NoStop}%
\bibitem [{Note2()}]{Note2}%
  \BibitemOpen
  \bibinfo {note} {See supplementary material for information on
  truncation.}\BibitemShut {Stop}%
\bibitem [{\citenamefont {Noponen}\ and\ \citenamefont
  {Turunen}(1994)}]{Noponen1994}%
  \BibitemOpen
  \bibfield  {author} {\bibinfo {author} {\bibfnamefont {E.}~\bibnamefont
  {Noponen}}\ and\ \bibinfo {author} {\bibfnamefont {J.}~\bibnamefont
  {Turunen}},\ }\bibfield  {title} {\bibinfo {title} {{Eigenmode method for
  electromagnetic synthesis of diffractive elements with three-dimensional
  profiles}},\ }\href {https://doi.org/10.1364/JOSAA.11.002494} {\bibfield
  {journal} {\bibinfo  {journal} {Journal of the Optical Society of America A}\
  }\textbf {\bibinfo {volume} {11}},\ \bibinfo {pages} {2494} (\bibinfo {year}
  {1994})}\BibitemShut {NoStop}%
\bibitem [{\citenamefont {Dietrich}\ \emph {et~al.}(2012)\citenamefont
  {Dietrich}, \citenamefont {Lehr}, \citenamefont {Helgert}, \citenamefont
  {T{\"{u}}nnermann},\ and\ \citenamefont {Kley}}]{Dietrich2012}%
  \BibitemOpen
  \bibfield  {author} {\bibinfo {author} {\bibfnamefont {K.}~\bibnamefont
  {Dietrich}}, \bibinfo {author} {\bibfnamefont {D.}~\bibnamefont {Lehr}},
  \bibinfo {author} {\bibfnamefont {C.}~\bibnamefont {Helgert}}, \bibinfo
  {author} {\bibfnamefont {A.}~\bibnamefont {T{\"{u}}nnermann}},\ and\ \bibinfo
  {author} {\bibfnamefont {E.-B.}\ \bibnamefont {Kley}},\ }\bibfield  {title}
  {\bibinfo {title} {{Circular Dichroism from Chiral Nanomaterial Fabricated by
  On-Edge Lithography}},\ }\href {https://doi.org/10.1002/adma.201203424}
  {\bibfield  {journal} {\bibinfo  {journal} {Advanced Materials}\ }\textbf
  {\bibinfo {volume} {24}},\ \bibinfo {pages} {OP321} (\bibinfo {year}
  {2012})}\BibitemShut {NoStop}%
\bibitem [{\citenamefont {Helgert}\ \emph {et~al.}(2011)\citenamefont
  {Helgert}, \citenamefont {Pshenay-Severin}, \citenamefont {Falkner},
  \citenamefont {Menzel}, \citenamefont {Rockstuhl}, \citenamefont {Kley},
  \citenamefont {T{\"{u}}nnermann}, \citenamefont {Lederer},\ and\
  \citenamefont {Pertsch}}]{Helgert2011}%
  \BibitemOpen
  \bibfield  {author} {\bibinfo {author} {\bibfnamefont {C.}~\bibnamefont
  {Helgert}}, \bibinfo {author} {\bibfnamefont {E.}~\bibnamefont
  {Pshenay-Severin}}, \bibinfo {author} {\bibfnamefont {M.}~\bibnamefont
  {Falkner}}, \bibinfo {author} {\bibfnamefont {C.}~\bibnamefont {Menzel}},
  \bibinfo {author} {\bibfnamefont {C.}~\bibnamefont {Rockstuhl}}, \bibinfo
  {author} {\bibfnamefont {E.-B.}\ \bibnamefont {Kley}}, \bibinfo {author}
  {\bibfnamefont {A.}~\bibnamefont {T{\"{u}}nnermann}}, \bibinfo {author}
  {\bibfnamefont {F.}~\bibnamefont {Lederer}},\ and\ \bibinfo {author}
  {\bibfnamefont {T.}~\bibnamefont {Pertsch}},\ }\bibfield  {title} {\bibinfo
  {title} {{Chiral Metamaterial Composed of Three-Dimensional Plasmonic
  Nanostructures}},\ }\href {https://doi.org/10.1021/nl202565e} {\bibfield
  {journal} {\bibinfo  {journal} {Nano Letters}\ }\textbf {\bibinfo {volume}
  {11}},\ \bibinfo {pages} {4400} (\bibinfo {year} {2011})}\BibitemShut
  {NoStop}%
\bibitem [{\citenamefont {Pshenay-Severin}\ \emph {et~al.}(2014)\citenamefont
  {Pshenay-Severin}, \citenamefont {Falkner}, \citenamefont {Helgert},\ and\
  \citenamefont {Pertsch}}]{Pshenay-Severin2014a}%
  \BibitemOpen
  \bibfield  {author} {\bibinfo {author} {\bibfnamefont {E.}~\bibnamefont
  {Pshenay-Severin}}, \bibinfo {author} {\bibfnamefont {M.}~\bibnamefont
  {Falkner}}, \bibinfo {author} {\bibfnamefont {C.}~\bibnamefont {Helgert}},\
  and\ \bibinfo {author} {\bibfnamefont {T.}~\bibnamefont {Pertsch}},\
  }\bibfield  {title} {\bibinfo {title} {{Ultra broadband phase measurements on
  nanostructured metasurfaces}},\ }\href {https://doi.org/10.1063/1.4881332}
  {\bibfield  {journal} {\bibinfo  {journal} {Applied Physics Letters}\
  }\textbf {\bibinfo {volume} {104}},\ \bibinfo {pages} {221906} (\bibinfo
  {year} {2014})}\BibitemShut {NoStop}%
\end{thebibliography}%

\end{document}